\documentclass[pdflatex,sn-nature]{sn-jnl}

\usepackage{graphicx}%
\usepackage{subcaption}
\usepackage{multirow}%
\usepackage{amsmath,amssymb,amsfonts}%
\usepackage{amsthm}%
\usepackage{mathrsfs}%
\usepackage[title]{appendix}%
\usepackage{xcolor}%
\usepackage{textcomp}%
\usepackage{manyfoot}%
\usepackage{booktabs}%
\usepackage{algorithm}%
\usepackage{algorithmicx}%
\usepackage{algpseudocode}%
\usepackage{listings}%
\usepackage{tabularx}%
\usepackage{bm}
\usepackage{hyperref}

\newcolumntype{R}[1]{>{\raggedleft\arraybackslash}p{#1}}

\theoremstyle{thmstyleone}%
\theoremstyle{thmstyletwo}%

\theoremstyle{thmstylethree}%

\begin{document}

\title[The role of Projects of Common Interest in reaching Europe's energy policy targets]{The role of Projects of Common Interest in reaching Europe's energy policy targets}


\author*[1]{\fnm{Bobby} \sur{Xiong}}\email{xiong@tu-berlin.de}

\author[1]{\fnm{Tom} \sur{Brown}}\email{t.brown@tu-berlin.de}

\author[1]{\fnm{Iegor} \sur{Riepin}}\email{iegor.riepin@tu-berlin.de}

\affil[1]{\orgdiv{Department of Digital Transformation in Energy Systems}, \orgname{Insitute of Energy Technology, Technische Universität Berlin}, \orgaddress{\city{Berlin}, \country{Germany}}}

\abstract{
    The European Union aims to achieve climate-neutrality by 2050, with interim 2030 targets including 55\% greenhouse gas emissions reduction compared to 1990 levels, 10 Mt p.a. of a domestic green H$_2$ production, and 50 Mt p.a. of domestic CO$_2$ injection capacity. To support these targets, Projects of Common and Mutual Interest (PCI-PMI) --- large infrastructure projects for electricity, hydrogen and CO$_2$ transport, and storage --- have been identified by the European Commission. This study focuses on PCI-PMI projects related to hydrogen and carbon value chains, assessing their long-term system value and the impact of pipeline delays and shifting policy targets using the sector-coupled energy system model PyPSA-Eur.
    Our study shows that PCI-PMI projects enable a more cost-effective transition to a net-zero energy system compared to scenarios without any pipeline expansion.
    Hydrogen pipelines help distribute affordable green hydrogen from renewable-rich regions in the north and southwest to high-demand areas in central Europe, while CO$_2$ pipelines link major industrial emitters with offshore storage sites. Although these projects are not essential in 2030, they begin to significantly reduce annual system costs by more than €26 billion from 2040 onward. Delaying implementation beyond 2040 could increase system costs by up to €24.2 billion per year, depending on the extent of additional infrastructure development. Moreover, our results show that PCI-PMI projects reduce the need for excess wind and solar capacity and lower reliance on individual CO$_2$ removal technologies, such as Direct Air Capture, by 13 to 136 Mt annually, depending on the build-out scenario.
}


\maketitle

With the European Green Deal, the European Union (EU) set a strategic path to become climate-neutral by 2050, with interim Greenhouse Gas (GHG) emission reduction targets of 55\% by 2030 compared to 1990 levels \cite{europeancommissionFit55Delivering2021}. Both the net-zero target and the interim 2030 goals are legally binding under the European Climate Law \cite{europeanparliamentRegulationEU20212021}. In practice, these policy targets mean transforming the EU into `a modern, resource-efficient and competitive' economy with net-zero GHG emissions \cite{europeancommissionEuropeanGreenDeal2021}. Current industrial processes and economic growth will need to be decoupled from fossil fuel dependencies. To achieve this transition across all sectors, the EU needs to scale up a portfolio of renewable energy sources, power-to-X solutions, Carbon Capture, Utilisation and Storage (CCUS), and Carbon Dioxide Removal (CDR) technologies, such as Direct Air Capture (DAC). In parallel, complementing investments into the electricity grid, hydrogen (H$_2$) and carbon dioxide (CO$_2$) transport and storage infrastructure are essential for efficient distribution across the European continent \cite{hofmannH2CO2Network2025}.

\paragraph{Hydrogen}
Hydrogen is expected to occupy a key position in this transition as it is considered essential for decarbonising hard-to-abate sectors, such as, but not limited to steel, refining, fertilisers, shipping, and aviation \cite{beresWillHydrogenSynthetic2024,neumannPotentialRoleHydrogen2023}. To lay out the foundation for a future hydrogen economy, the EU has set ambitious targets for domestic hydrogen production and infrastructure build-out. Under the EU Hydrogen Strategy \cite{europeancommissionCommunicationCommissionEuropean2020}, reinforced by REPowerEU \cite{europeancommissionREPowerEUPlanCommunication2022} and the Net-Zero Industry Act (NZIA) \cite{europeanparliamentRegulationEU20242024}, the EU aims to install at least 40 GW electrolysis capacity by 2030, domestically (with an additional 40 GW to be installed in so-called European Neighbourhood countries \cite{europeanparliamentRegulationEU20212021a}). REPowerEU foresees the annual production of 10 Mt of domestic renewable hydrogen by 2030, alongside an additional 10 Mt sourced through imports \cite{europeancommissionREPowerEUPlanCommunication2022}. Initiatives like the European Hydrogen Backbone (EHB) aim to support this transition by proposing a hydrogen transport network across Europe. The EHB initiative envisions a H$_2$ pipeline network of almost 53000 km by 2040 \cite{europeanhydrogenbackboneinitiativeEuropeanHydrogenBackbone2022}, including repurposing existing natural gas infrastructure and new potential routes.

\paragraph{CCUS}
Complementing its hydrogen ambitions, the EU has proposed similarly strategic plans for the carbon economy. In the Industrial Carbon Management Strategy, the EU envisages a single market for CO$_2$ in Europe, to enable CO$_2$ to become a tradable commodity for storage, sequestration, or utilisation \cite{europeancourtofauditorsEUsIndustrialPolicy2024}. Beyond a net-zero emission target in the European Climate Law \cite{europeanparliamentRegulationEU20212021}, CO$_2$ serves as a key feedstock for the production of synthetic fuels, such as methanol, methane, as well as high-value chemicals \cite{neumannPotentialRoleHydrogen2023}. Outside of CO$_2$ utilisation, Carbon Capture and Storage (CCS) is considered indispensable for achieving net-zero emissions in sectors with unavoidable process-based CO$_2$ emissions, such as cement, chemicals, and waste-to-energy. Here, the NZIA mandates that all EU member states collectively ensure that at least 50 Mt p.a. of CO$_2$ can be injected and stored by 2030. The European Commission further estimates that up to 550 Mt p.a. of CO$_2$ will need to be captured by 2050 \cite{europeanparliamentRegulationEU20242024}. At least 250 Mt p.a. will need to be sequestered in the European Economic Area \cite{europeancommissionCommunicationCommissionEuropean2024}.

\begin{table}[htbp]
  \centering
  \caption{Literature overview of studies modelling CO$_2$ and H$_2$ in low-carbon energy systems.}
  \label{tab:literature}
  \scriptsize
  \begin{tabularx}{\linewidth}{R{1.3cm}>{\hsize=2\hsize\centering\arraybackslash}X>{\hsize=2\hsize\centering\arraybackslash}X>{\hsize=6\hsize\centering\arraybackslash}X>{\hsize=20\hsize\centering\arraybackslash}X}
    \toprule
    & \textbf{CO$_2$} & \textbf{H$_2$} & \textbf{Region} & \textbf{Methodology} \\
    \midrule
    \textbf{Study} & & & & \\
    \cite{vangreevenbroekLittleLoseCase2025} & $\square$ & $\square$ & Europe & myopic (2025-2050), PyPSA-Eur\\
    \cite{hofmannH2CO2Network2025} & $\blacksquare$ & $\blacksquare$ & Europe & overnight (2050), PyPSA-Eur\\
    \cite{kountourisUnifiedEuropeanHydrogen2024} & $\square$  & $\blacksquare$ & Europe & myopic (2030-2050), Balmorel \cite{wieseBalmorelOpenSource2018} \\
    \cite{neumannPotentialRoleHydrogen2023} & $\square$ & $\blacksquare$ & Europe & overnight (2050), PyPSA-Eur \\
    \cite{neumannGreenEnergySteel2025} & $\square$ & $\blacksquare$ & Europe & overnight (2050), PyPSA-Eur, TRACE \cite{hamppImportOptionsChemical2023} \\
    \cite{beresWillHydrogenSynthetic2024} & -- & $\square$ & Europe & overnight (2050), JRC-EU-TIMES \\
    \cite{fleiterHydrogenInfrastructureFuture2025} & $\square$ & $\blacksquare$ & Europe & overnight (2030, 2050), Enertile \\
    \cite{cerniauskasOptionsNaturalGas2020} & -- & $\blacksquare$ & Germany & overnight (2030), supply chain model \cite{reussSeasonalStorageAlternative2017} \\
    \cite{bakkenLinearModelsOptimization2008} & $\square$ & -- & Norway & myopic (2010-2030), investment model \\
    \bottomrule
  \end{tabularx}
  \centering
  \footnotetext{$\blacksquare$ considered, including pipelines \quad  $\square$ considered, copperplated \quad -- not considered}
\end{table}

A growing body of literature has been investigating the long-term role of H$_2$ and CO$_2$ in low-carbon or net-zero energy systems (Table \ref{tab:literature}). Both carriers see their primary value outside the electricity sector, i.e., in the decarbonisation of hard-to-abate sectors such as industry, transport, shipping, and aviation \cite{reigstadMovingLowcarbonHydrogen2022,vangreevenbroekLittleLoseCase2025}. While there are direct use cases for H$_2$ in the industry sector such as steel production, it is primarily expected to serve as a precursor for synthetic fuels, including methanol, Fischer-Tropsch fuels (e.g. synthetic kerosene and naphta) and methane \cite{neumannNearoptimalFeasibleSpace2021,neumannPotentialRoleHydrogen2023,neumannGreenEnergySteel2025,kountourisUnifiedEuropeanHydrogen2024,beresWillHydrogenSynthetic2024,fleiterHydrogenInfrastructureFuture2025}. To produce these carbonaceous fuels, CO$_2$ is required as a feedstock (Carbon Utilisation --- CU). This CO$_2$ can be captured from the atmosphere via DAC, biomass plants, or from industrial and process emissions (e.g. cement, steel, ammonia production) in combination with Carbon Capture (CC) units \cite{hofmannH2CO2Network2025,bakkenLinearModelsOptimization2008}.

\paragraph{Transport infrastructure and PCI-PMI projects}
To meet the need for green electricity, green H$_2$ and CO$_2$, significant investments into its transport and storage/sequestration infrastructure are needed \cite{kountourisUnifiedEuropeanHydrogen2024,neumannPotentialRoleHydrogen2023,cerniauskasOptionsNaturalGas2020}. A recent report by the European Commission confirms that investment needs into the EU's energy infrastructure will continue to grow \cite{europeancommission.directorategeneralforenergy.InvestmentNeedsEuropean2025}, estimating planned expenditures of around €170 billion for H$_2$ and up to €20 billion for CO$_2$ infrastructure by 2040, respectively. It also emphasises that these investments face higher uncertainty, as both sectors are still in their infancy. 

\begin{figure}[htbp]
  \centering
  \includegraphics{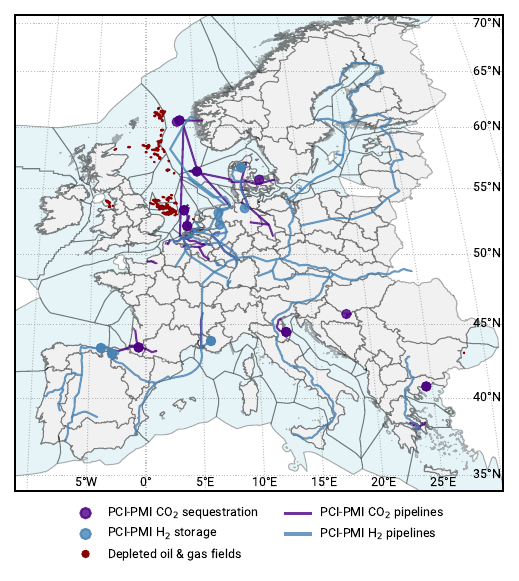}
  \caption{Map of the regional scope including clustered onshore (grey) and offshore regions (blue), as well as PCI-PMI CO$_2$ and H$_2$ pipelines, storage and sequestration sites. Depleted offshore oil and gas fields (red) provide additional CO$_2$ sequestration potential \cite{hofmannH2CO2Network2025}.}
  \label{fig:regional_scope_map}
\end{figure}

Within the transition towards net-zero, the EU has established a framework to support the development of key cross-border and national infrastructure projects, which are considered essential for achieving the EU's energy policy targets. These Projects of Common Interest (PCI) are projects that link the energy systems of two or more EU member states \cite{europeancommissionRegulationEUNo2022}. In a biennial selection process, PCIs are identified through regional stakeholder groups and evaluated based on their contribution to the EU's energy security, e.g. by improving market integration, diversification of energy supply, and integration of renewables. So-called Projects of Mutual Interest (PMI) transfer the same concept to projects that link the EU's energy system with third countries, such as Norway or the United Kingdom, the Western Balkans or North Africa, as long as they align with EU climate and energy objectives \cite{europeancommissionCommissionDelegatedRegulation2023}. Approved PCI-PMI projects benefit from accelerated permitting and access to EU funding under the Connecting Europe Facility (CEF). Given the strong political and project promoter support, comprehensive reporting and monitoring processes, as well as their role as technological lighthouses, projects on the PCI-PMI list are more likely to be implemented than others \cite{europeancommission.directorategeneralforenergy.InvestmentNeedsEuropean2025}. Nonetheless, large infrastructure projects—including those on the PCI-PMI list—often face delays due to permitting hurdles, financing constraints, procurement bottlenecks, and other implementation challenges \cite{acerConsolidatedReportProgress2023}. 
As a direct result of the revised TEN-E Regulation (EU 2022/869) \cite{europeanparliamentRegulationEU20222022}, the 2023 PCI-PMI list \cite{europeancommissionCommissionDelegatedRegulation2023,europeancommissionPCIPMITransparencyPlatform2024} for the first time includes H$_2$ and CO$_2$ transport and storage projects, alongside electricity and gas projects. A continent-wide hydrogen backbone --- connecting regions rich in renewable energy potential to industrial and storage hubs --- is viewed essential for transporting H$_2$ where it is needed. Likewise, CO$_2$ pipelines and sequestration sites are needed to capture, transport and sequester emissions from industrial processes and power plants. An overview of the PCI-PMI projects is provided in Figure \ref{fig:regional_scope_map}. With around 14 projects in the priority thematic area `cross-border carbon dioxide network' and 32 projects listed in `hydrogen interconnections' (including pipelines and electrolysers), this PCI-PMI list lays the foundation for a future pan-European H$_2$ and CO$_2$ value chain \cite{europeancommissionAnnexFirstUnion2023}.

\paragraph{Research gaps and contribution of this study}
Several studies (Table \ref{tab:literature}) have begun to explore the interaction between CO$_2$ and H$_2$ infrastructure in sector-coupled energy system models, however important aspects remain insufficiently addressed --- in particular the role of real planned infrastructure projects, transformation pathways, and the influence of uncertainties on the long-term performance of these projects. Many studies have demonstrated how uncertainty in future developments can significantly impact the cost-effectiveness of energy system investments \cite{vanderweijdeEconomicsPlanningElectricity2012,mobiusRegretAnalysisInvestment2020,neumannNearoptimalFeasibleSpace2021,vangreevenbroekLittleLoseCase2025,priceModellingGenerateAlternatives2017,yueReviewApproachesUncertainty2018}. Existing analyses abstract away from actual investment plans, such as whose under the PCI-PMI framework, potentially neglecting infrastructure options that are not perfectly cost-optimal but have a high likelihood of implementation, e.g., due to political support \cite{vangreevenbroekLittleLoseCase2025,trutnevyteDoesCostOptimization2016}. 
While Hofmann et al. \cite{hofmannH2CO2Network2025} provide valuable insights into the synergies between H$_2$ and CO$_2$ infrastructure, the lack of inclusion of planned projects and focus on a single modelling year might yield overly optimistic results. To our knowledge, the contribution of PCI-PMI projects has not yet been evaluated within a sector-coupled modelling framework that incorporates future policy targets, uncertainty and transformation pathways.

Our study addresses these gaps by explicitly including PCI-PMI projects in PyPSA-Eur \cite{horschPyPSAEurOpenOptimisation2018}, a sector-coupled model of the European energy system. We assess various build-out levels of CO$_2$ and H$_2$ infrastructure across short-term scenarios and transformation pathways. Using a myopic, iterative modelling approach, we simulate energy system development from 2030 to 2050 under non-anticipative foresight, reflecting the reality that market participants do not have perfect knowledge of long-term developments. This approach helps avoid the overly optimistic outcomes of long-term perfect foresight models. To quantify the economic value associated with PCI-PMI projects across scenarios reflecting a selected set of uncertainties --- including changes in EU energy policy project delays, and cancellations --- we use a regret-based approach (see \nameref{sec:scenarios}) By limiting the analysis to a set of scenarios, this regret analysis is manageable and computationally feasible.

This study also aims to reduce the uncertainty surrounding the `chicken-and-egg' dilemma in infrastructure investment --- whether to develop CO$_2$ and H$_2$ infrastructure in advance or to wait for demand to materialise. Specifically, we address the following research questions: 
\begin{enumerate} 
  \item What is the long-term value of PCI-PMI projects in supporting the EU’s climate and energy policy targets, and what are the associated costs?
  \item What are the costs of adhering to the EU policy targets, even when the implementation of PCI-PMI projects is delayed?
\end{enumerate}

\section*{Scenarios}\label{sec:scenarios}
\paragraph{Long-term scenarios vary in infrastructure build-out}
We define the long-term scenarios based on the degree of CO$_2$ and H$_2$ infrastructure build-out, including the roll-out of PCI-PMI projects as well additional pipeline investments. In total, we implement five long-term scenarios, (i) a pessimistic scenario (Decentral Islands --- \textit{DI}) without any H$_2$ pipeline and onshore CO$_2$ pipeline infrastructure, (ii) a scenario that considers the on-time commissioning of all PCI-PMI CO$_2$ and H$_2$ projects (PCI-PMI --- \textit{PCI}) only, (iii) more ambitious scenarios that further allow investments into national and (iv) international pipelines (PCI-PMI nat. --- \textit{PCI-n} and PCI-PMI internat. --- \textit{PCI-in}), and (v) a scenario that does not assume any fixed PCI-PMI infrastructure but allows for a centralised, purely needs-based build-out of CO$_2$ and H$_2$ pipelines (Centralised Planning --- \textit{CP}). An overview of the long-term scenarios and their associated model-endogenous decision variables is provided in Table \ref{tab:long-term_scenarios}. 

\begin{table}[htpb]
  \centering
  \caption{Overview of long-term scenarios and their key assumptions.}
  \label{tab:long-term_scenarios}
  \scriptsize
  \begin{tabularx}{\linewidth}{R{4.2cm}>{\centering\arraybackslash}X>{\centering\arraybackslash}X>{\centering\arraybackslash}X>{\centering\arraybackslash}X>{\centering\arraybackslash}X}
    \toprule
    \textbf{Long-term scenarios} & 
    \textbf{DI} & 
    \textbf{PCI} & 
    \textbf{PCI-n} & 
    \textbf{PCI-in} & 
    \textbf{CP} \\
    \midrule
    \textbf{CO$_2$ sequestration} & & & & & \\
    Depleted oil \& gas fields\footnotemark[1] & $\blacksquare$ & $\blacksquare$ & $\blacksquare$ & $\blacksquare$ & $\blacksquare$ \\
    PCI-PMI seq. sites\footnotemark[2] & -- & $\blacksquare$ & $\blacksquare$ & $\blacksquare$ & $\blacksquare$ \\
    \midrule
    \textbf{H$_2$ storage} & & & & & \\
    Endogenous build-out & $\blacksquare$ & $\blacksquare$ & $\blacksquare$ & $\blacksquare$ & $\blacksquare$ \\
    PCI-PMI storage sites & -- & $\blacksquare$ & $\blacksquare$ & $\blacksquare$ & $\blacksquare$ \\
    \midrule
    \textbf{CO$_2$ pipelines} & & & & & \\
    to depleted oil \& gas fields & $\blacksquare$ & $\blacksquare$ & $\blacksquare$ & $\blacksquare$ & $\blacksquare$ \\
    to PCI-PMI seq. sites & -- & $\blacksquare$ & $\blacksquare$ & $\blacksquare$ & $\blacksquare$ \\
    \midrule
    \textbf{CO$_2$ and H$_2$ pipelines} & & & & & \\
    PCI-PMI & -- & $\blacksquare$ & $\blacksquare$ & $\blacksquare$ & $\blacksquare$ \\
    National build-out & -- & $\blacksquare$ & $\blacksquare$ & $\blacksquare$ & $\blacksquare$ \\
    International build-out & -- & -- & -- & $\blacksquare$ & $\blacksquare$ \\
    PCI-PMI extendable & -- & -- & -- & -- & $\blacksquare$ \\
    \bottomrule
  \end{tabularx}
  \centering
  \footnotetext{$\blacksquare$ enabled\quad-- disabled}
  \footnotetext[1]{approx. 286 Mt p.a.}
  \footnotetext[2]{approx. 114 Mt p.a.}
\end{table}

\paragraph{Regret analysis to evaluate the impact of policy targets and delays}
In a subsequent step, we assess how different short-term scenarios impact the long-term decarbonisation pathways using a regret-based approach. In decision theory \cite{loomesRegretTheoryAlternative1982}, the concept of regret is typically defined as the difference in economic value, payoff, or cost between a chosen strategy and the optimal strategy under identical conditions \cite{mobiusRegretAnalysisInvestment2020}. The regret term then represents the additional cost incurred from not following the cost-optimal strategy. In energy modelling literature \cite{vanderweijdeEconomicsPlanningElectricity2012,mobiusRegretAnalysisInvestment2020}, a regret analysis is usually designed in two steps: First, a set of scenarios is defined, representing different future developments (see long-term scenarios above). In a second step, the performance of first-stage investment is evaluated under the realisation of second-stage or short-term realisations of the future \cite{salvatoreManagerialEconomicPrinciples2008}. It is particularly useful in energy system modelling, where future uncertainties can significantly impact the performance of investments in infrastructure and technologies. 

\begin{table}[htbp]
  \centering
  \caption{Regret matrix setup: Long-term and short-term scenarios.}
  \label{tab:regret_matrix_setup}
  \scriptsize
  \begin{tabularx}{\linewidth}{R{4.2cm}>{\centering\arraybackslash}X>{\centering\arraybackslash}X>{\centering\arraybackslash}X}
    \toprule
    \textbf{Short-term scenarios} & \textbf{Reduced targets} & \textbf{Delayed pipelines} & \textbf{No pipelines} \\
    \midrule
    \textbf{Long-term scenarios} & & & \\
    Decentral Islands (\textbf{DI}) & $\blacksquare$ & -- & -- \\
    PCI-PMI (\textbf{PCI}) & $\blacksquare$ & $\blacksquare$ & $\blacksquare$ \\
    PCI-PMI nat. (\textbf{PCI-n}) & $\blacksquare$ & $\blacksquare$ & $\blacksquare$\\
    PCI-PMI internat. (\textbf{PCI-in}) & $\blacksquare$ & $\blacksquare$ & $\blacksquare$ \\
    Central Planning (\textbf{CP}) & $\blacksquare$ & $\blacksquare$ & $\blacksquare$ \\
    \midrule
    \textbf{Targets} & & & \\
    GHG emission reduction &  $\blacksquare$ &  $\blacksquare$ &  $\blacksquare$ \\
    CO$_2$ sequestration &  -- &  $\blacksquare$ &  $\blacksquare$ \\
    Electrolytic H$_2$ production &  -- &  $\blacksquare$ &  $\blacksquare$ \\
    H$_2$ electrolysers &  -- &  $\blacksquare$ &  $\blacksquare$ \\
    \midrule
    \textbf{CO$_2$ + H$_2$ infrastructure} & & & \\
    CO$_2$ sequestration sites & $\blacksquare$ &  $\blacksquare$ &  $\blacksquare$ \\
    CO$_2$ pipelines to seq. site & $\blacksquare$ &  $\blacksquare$ &  $\blacksquare$ \\
    CO$_2$ pipelines & $\blacksquare$ &  $\square$ &  -- \\
    H$_2$ pipelines & $\blacksquare$ &  $\square$ &  -- \\
    \bottomrule
  \end{tabularx}
  \centering
  \footnotetext{$\blacksquare$ enabled, $\square$ delayed by one period, -- disabled}
  \footnotetext{Considering the total number of long-term (5) and short-term scenarios (3) as well as planning horizons (3), the regret matrix yields 60 optimisation problems (15+45).}
\end{table}

Table \ref{tab:regret_matrix_setup} gives an overview of the regret matrix setup and its underlying assumptions, where the long-term scenario serves as the \textit{planned} or \textit{anticipated} and the short-term scenario serves as the hypothetically \textit{realised} outcome. Specifically, we assume that the CO$_2$ and H$_2$ pipeline capacities identified in the long-term modelling exercise are either maintained at their planned levels, delayed in implementation, or not built at all.
In the short-term scenarios, the model can still react by investing into additional generation, storage, or conversion, or carbon-removal technologies, assuming the technical potential was not exceeded in the long-term optimisation. At this step, we also simulate changes in energy policy. For example, in \textit{Reduced targets}, we remove all of the long-term targets (Table \ref{tab:targets}) except for the GHG emission reduction targets to assess the value of the CO$_2$ and H$_2$ infrastructure in a less ambitious policy environment \cite{europeancourtofauditorsEUsIndustrialPolicy2024}. In \textit{Delayed pipelines}, we assume that all PCI-PMI and endogenous pipelines are delayed by one period, i.e., the commissioning of the project is shifted to the next planning horizon. Lastly, we remove all pipeline capacities in \textit{No pipelines}, including the PCI-PMI projects, allowing us to evaluate the impact of a complete lack of planned infrastructure. 

\section*{Results}\label{sec:results}

\begin{figure}[htbp]
  \centering
  \includegraphics{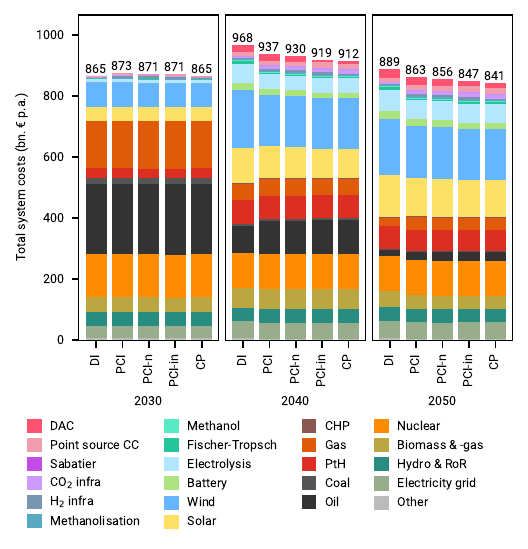}
  \caption{Total annual system costs (CAPEX + OPEX) by technology group. \textit{CO$_2$ and H$_2$ infrastructure each include pipelines, storage and sequestration sites, respectively. Gas refers to gas power plants and boilers. Coal infrastructure refers to hard coal and lignite power plants. Other includes SMR, rural heat, and thermal storage.}}
  \label{fig:costs_overview}
\end{figure}

\paragraph{Investing into PCI-PMI infrastructure unlocks significant cost savings}
Figure \ref{fig:costs_overview} shows the total annual system costs --- distributed over all modelled technology groups --- for each planning horizon and long-term scenario. We observe the highest total annual system costs in the planning horizon 2040, ranging from €912 to €968 billion per year. This cost increase is primarily driven by the sharp decarbonisation pathway planned for 2030 to 2040 --- a carbon budget reduction of more than 1600 Mt p.a. compared to the remaining 460 Mt p.a. in the last decade from 2040 to 2050. In 2030, total system costs are lowest in the \textit{DI} and \textit{CP} scenario, as the model does not see the need for large-scale investments into H$_2$ and CO$_2$ infrastructure yet (due to myopic foresight). Adding PCI-PMI projects in 2030 increases costs by less than 1\% (Figure \ref{fig:costs_overview}). With CO$_2$ pipelines connecting depleted offshore oil and gas fields to their closest onshore region, the policy targets, including CO$_2$ sequestration can be achieved at a total of €865 billion per year.

\begin{figure}[htbp]
  \centering
  \includegraphics{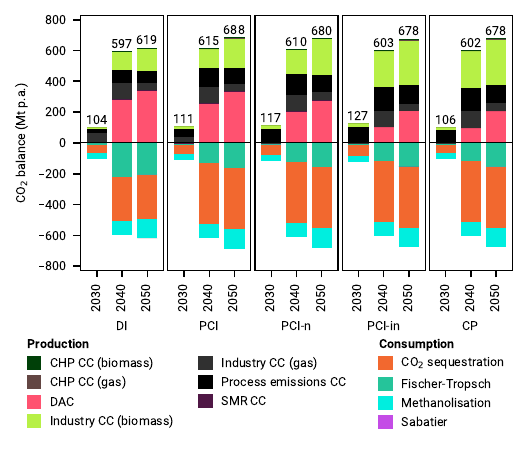}
  \caption{CO$_2$ balances in long-term scenarios.}
  \label{fig:balances_overview_co2_stored}
\end{figure}

\begin{figure}[htbp]
  \centering
  \includegraphics{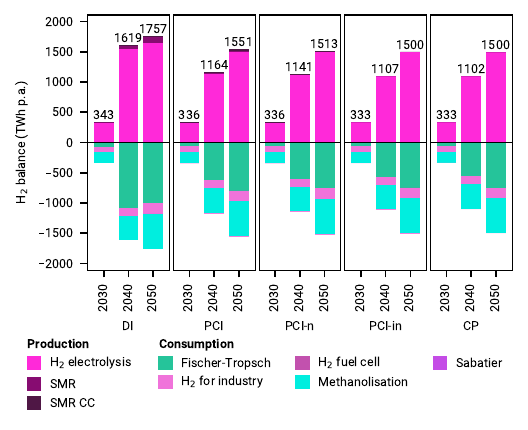}
  \caption{H$_2$ balances in long-term scenarios.}
  \label{fig:balances_overview_H2}
\end{figure}

Starting in 2040, all scenarios with PCI-PMI and endogenous pipeline investments unlock significant cost savings, from more than €30 billion per year in the \textit{PCI} up to €50 billion per year in the \textit{PCI-in} scenario, where additional pipeline build-out is allowed (see \nameref{sec:scenarios} section).
By granting the model complete flexibility to expand hydrogen and CO$_2$ infrastructure at any location beyond the PCI-PMI projects, we unlock additional annual cost savings of €6 to €7 billion per year through investments in fewer, yet more optimally located CO$_2$ and H$_2$ pipelines from a systemic perspective (see \textit{PCI-in} pipeline utilisation in Figure \ref{fig:PCI-in_lt} compared to \textit{CP} pipeline utilisation in Figure \ref{fig:CP_lt}).
Further, this reduces the reliance on larger investments into wind generation and costly DAC technologies near the sequestration sites. These effects are slightly less pronounced in the 2050 model results, where system costs can be reduced by €26 to €41 billion per year with PCI-PMI and endogenous pipeline investments. Here, higher Carbon Capture and Utilisation (CCU) via methanol synthesis and Fischer-Tropsch processes, supported by increased H$_2$ production as a chemical feedstock, enhances system flexibility compared to 2040 (Figures \ref{fig:balances_overview_co2_stored} and \ref{fig:balances_overview_H2}).

\begin{figure*}[htbp]
  \centering
  \begin{subfigure}[t]{0.32\textwidth}
      \vspace{0pt}
      \includegraphics[width=1\textwidth]{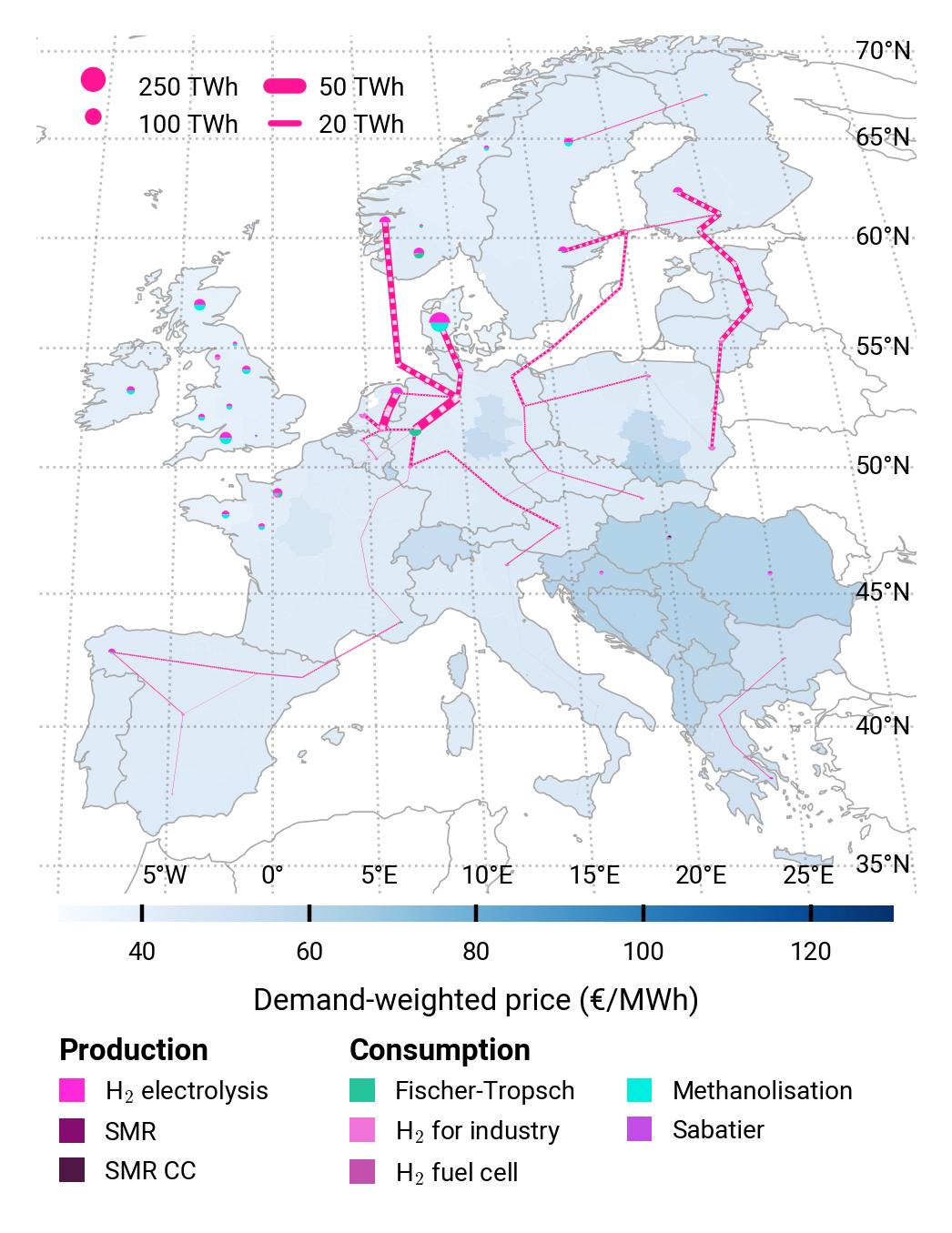}
      \caption{H$_2$ 2030.}
      \label{fig:PCI_lt_2030_h2}
  \end{subfigure}
  \begin{subfigure}[t]{0.32\textwidth}
      \vspace{0pt}
      \includegraphics[width=1\textwidth]{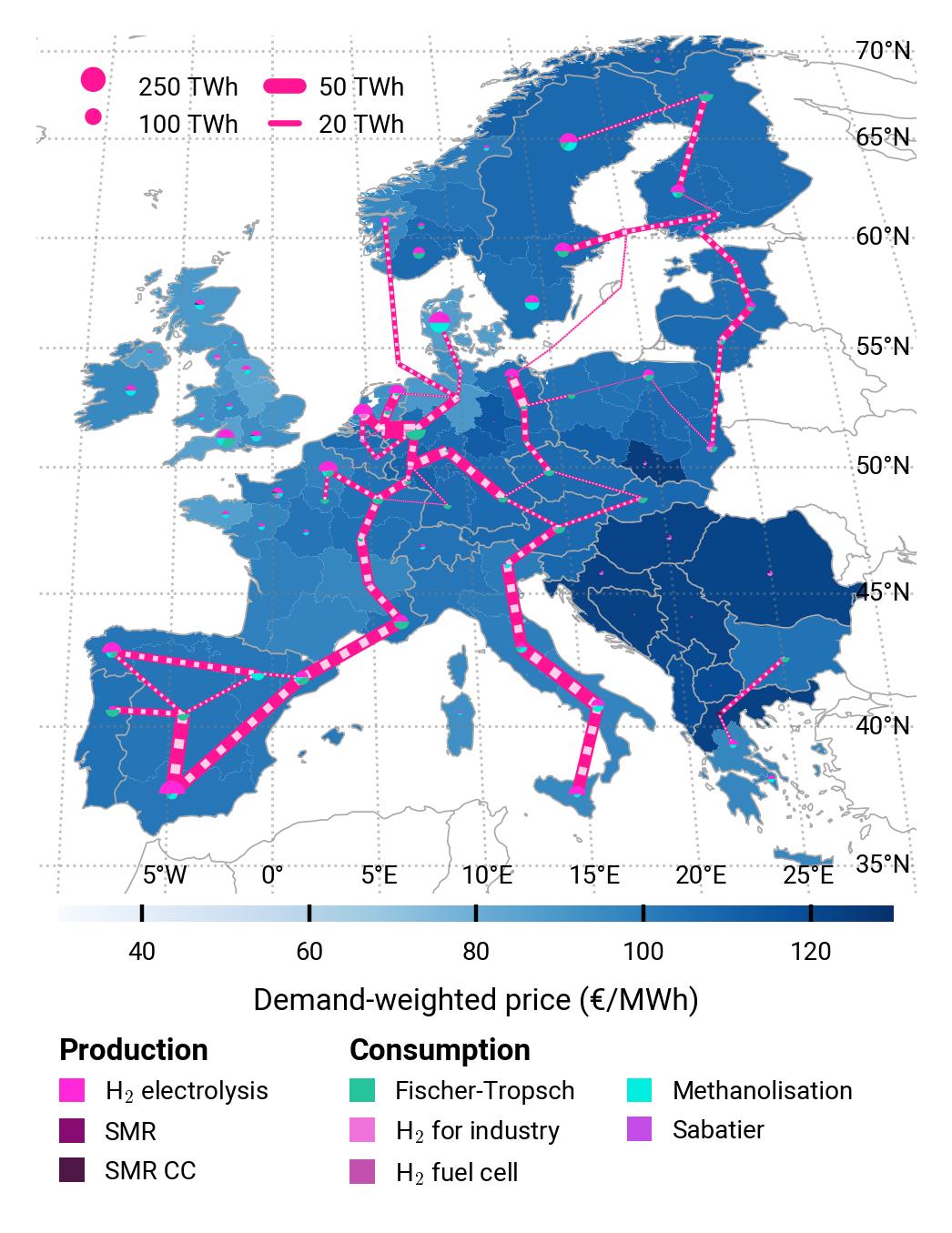}
      \caption{H$_2$ 2040.}
      \label{fig:PCI_lt_2040_h2}
  \end{subfigure}
  \begin{subfigure}[t]{0.32\textwidth}
    \vspace{0pt}
    \includegraphics[width=1\textwidth]{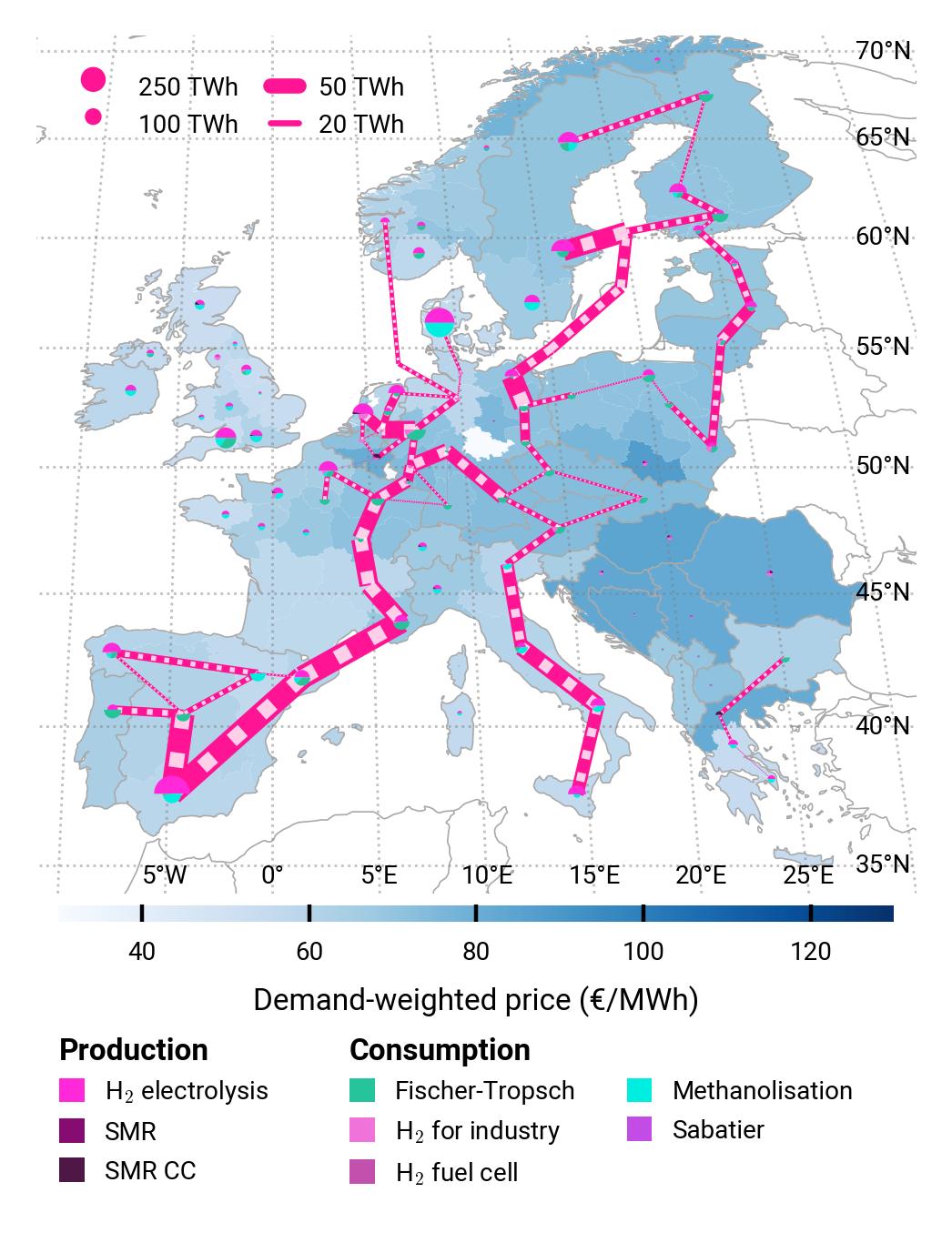}
    \caption{H$_2$ 2050.}
    \label{fig:PCI_lt_2050_h2}
  \end{subfigure}
  \begin{subfigure}[t]{0.32\textwidth}
      \vspace{0pt}
      \includegraphics[width=1\textwidth]{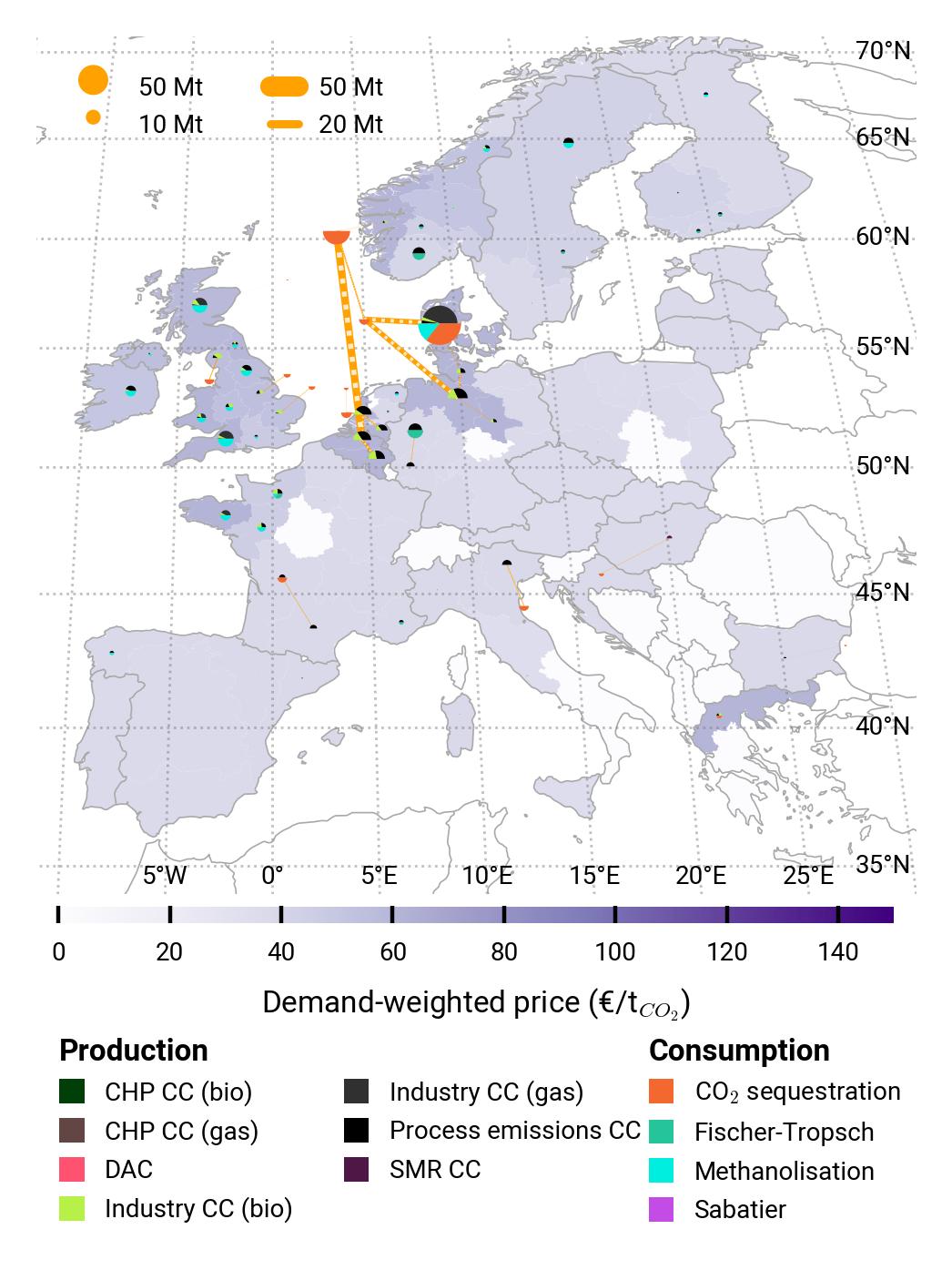} 
      \caption{CO$_2$ 2030.}
      \label{fig:PCI_lt_2030_co2}
  \end{subfigure}
  \begin{subfigure}[t]{0.32\textwidth}
      \vspace{0pt}
      \includegraphics[width=1\textwidth]{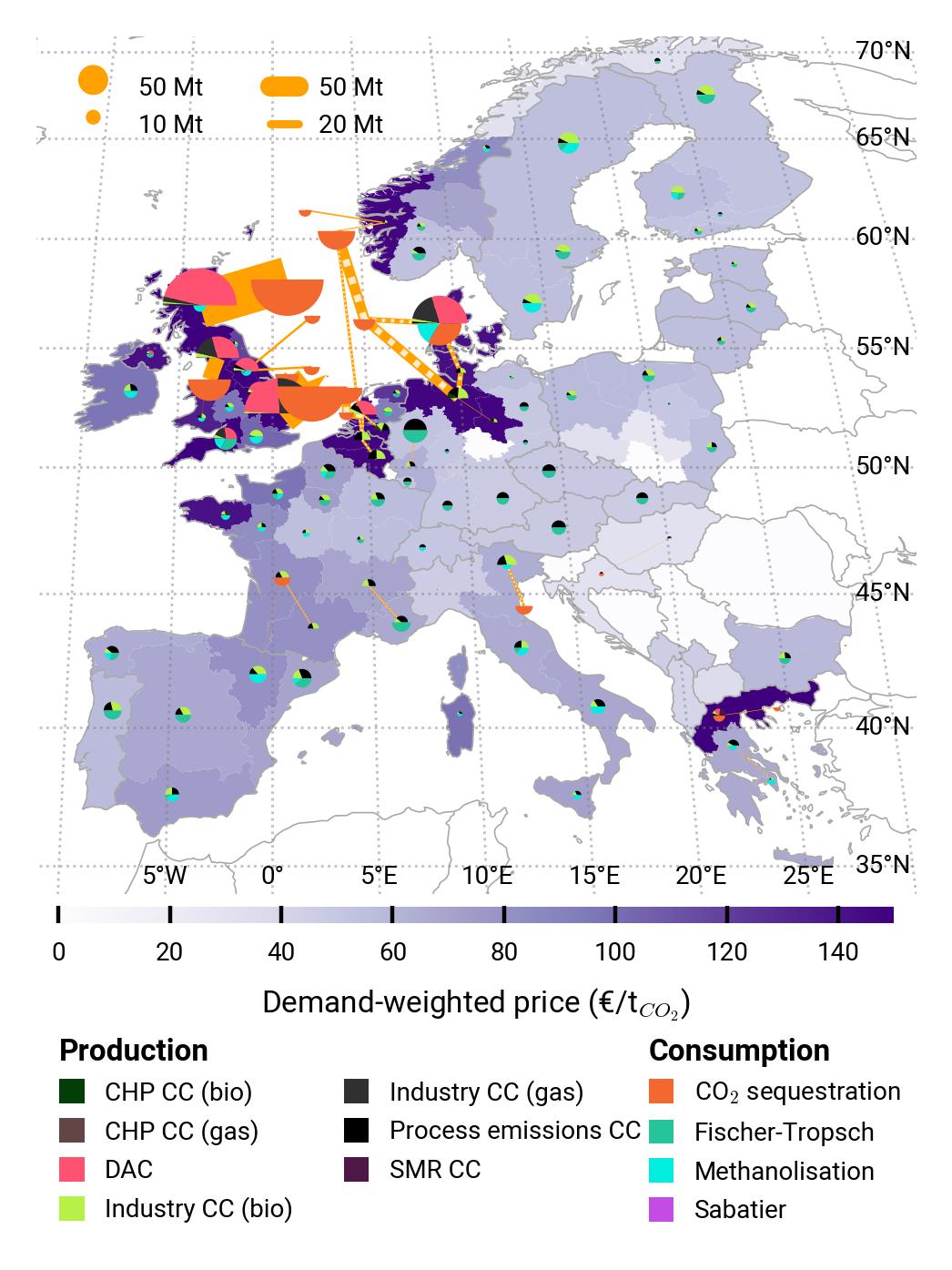} 
      \caption{CO$_2$ 2040.}
      \label{fig:PCI_lt_2040_co2}
  \end{subfigure}
  \begin{subfigure}[t]{0.32\textwidth}
      \vspace{0pt}
      \includegraphics[width=1\textwidth]{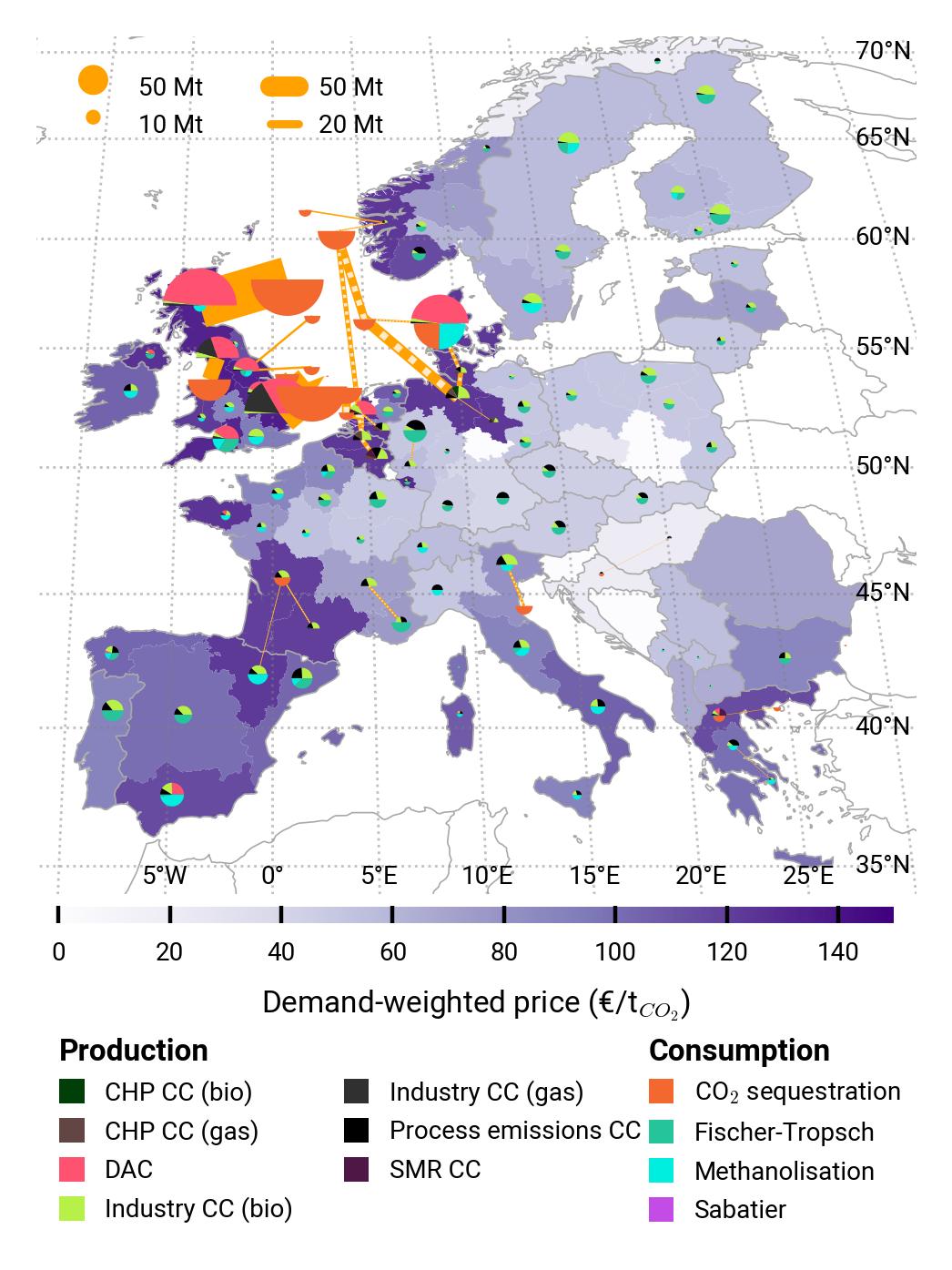} 
      \caption{CO$_2$ 2050.}
      \label{fig:PCI_lt_2050_co2}
  \end{subfigure}
  \vspace{0.3cm}
  \caption{\textit{PCI-PMI} long-term scenario --- Regional distribution of H$_2$ and CO$_2$ production, utilisation, storage, transport and price. Note that both the H$_2$ and CO$_2$ price refer to their value as a commodity, i.e., price is higher where there is a demand for it.}
  \label{fig:PCI_lt}
\end{figure*}

\paragraph{Carbon Capture, Utilisation, and Storage}
We find that most of the differences in system cost and savings can be attributed to the production and utilisation of CO$_2$, as shown in Figure \ref{fig:balances_overview_co2_stored}. Lacking the option to transport CO$_2$ from industry and other point sources to the offshore sequestration sites, the system requires expensive DAC in the \textit{DI} scenario. While the sequestration target of 50 Mt p.a. in 2030 is binding only in the \textit{DI} scenario, all other scenarios achieve higher levels of CO$_2$ sequestration as their CO$_2$ pipeline build-out increases. 
The 53.9 Mt p.a. of CO$_2$ sequestered in the \textit{CP} scenario serves as an indicator of the cost-optimal level of sequestration for the European energy system in 2030 assuming perfectly located pipeline infrastructure. With the inclusion of PCI-PMI projects, CO$_2$ sequestration ranges from 58.7 Mt p.a. in the \textit{PCI} to 75 Mt p.a. in the \textit{PCI-in} scenario. 
Looking at 2040 and 2050, in place of expensive DAC in the \textit{DI} scenario, the model equips biomass-based industrial processes --- primarily located in Belgium, the Netherlands and Western regions of Germany --- with carbon capture (see Figures \ref{fig:PCI_lt_2030_co2}, \ref{fig:PCI_lt_2040_co2}, and \ref{fig:PCI_lt_2050_co2}). 

In 2040 and 2050, all sequestration targets (Table \ref{tab:targets}) are overachieved, as the full combined CO$_2$ sequestration potential of 398 Mt p.a. is exploited in all scenarios where PCI-PMI projects are included (\textit{PCI} to \textit{CP}).  Emissions are captured from industrial processes equipped with carbon capture units, with biomass-based industry contributing the largest share of point-source carbon capture. This ranges from 119 to 241 Mt p.a. in 2040 and from 149 to 287 Mt p.a. in 2050, increasing with the build-out of CO$_2$ infrastructure (from left to right; see Figure 3). As the most expensive carbon capture option, CO$_2$ capture from SMR CC processes is limited to a maximum of 8 Mt p.a. in the \textit{PCI} scenario by 2050.
With a lower sequestration potential of 286 Mt p.a. in \textit{DI} scenario, more CO$_2$ is used as a precursor for the synthesis of Fischer-Tropsch fuels instead --- 221 Mt p.a. vs. 115-127 Mt p.a. in 2040 and 206 Mt p.a. vs. 153-163 Mt p.a. in 2050, to meet the emission reduction targets for 2040 and 2050, respectively. 
Given the fixed exogenous demand for shipping methanol (Figure \ref{fig:exogenous_demand}), CO$_2$ demand for methanolisation is constant across all scenarios (39 Mt p.a. in 2030, 89 Mt p.a. in 2040, and 127 Mt p.a. in 2050). 

\paragraph{Hydrogen production and utilisation}
H$_2$ production in the model is primarily driven by the demand for Fischer-Tropsch fuels and methanol. In 2030 and 2050, the electrolytic H$_2$ production target of 10 and 45 Mt p.a. is binding, equivalent to 333 and 1500 TWh p.a. (at a lower heating value of 33.33 kWh/kg for H$_2$). Only in 2040, the H$_2$ production target of 27.5 Mt p.a. (917 TWh p.a.) is overachieved by 185-247 TWh p.a. in the \textit{PCI} to \textit{CP} scenarios. H$_2$ production in the \textit{DI} is significantly higher, given its need for additional Fischer-Tropsch synthesis to bind CO$_2$ as an alternative to sequestration, as described in the previous section.
In 2050, Fischer-Tropsch fuels are primarily used to satisfy the demand for kerosene in aviation and naphta for industrial processes (see Table \ref{fig:exogenous_demand}). Only about 93 to 173 TWh p.a. of hydrogen is directly used in the industrial sector. Across all long-term scenarios, hydrogen is almost exclusively produced via electrolysis. Note that the model includes a green hydrogen production constraint reflecting energy policy targets, though it does not enforce an hourly matching rule.
In the \textit{DI} scenario, where there is no hydrogen pipeline infrastructure, the model resorts to Steam Methane Reforming (SMR) to produce 71 to 102 TWh p.a. of hydrogen in 2040 and 2050, respectively.

Geographically, H$_2$ production is concentrated in regions with high solar PV potential such as the Iberian and Italian Peninsula, as well as high wind infeed regions including Denmark, the Netherlands and Belgium. The produced H$_2$ is then transported via H$_2$ pipelines including PCI-PMI projects to carbon point sources  in central, continental Europe where it is used as a precursor for Fischer-Tropsch fuels. Onsite H$_2$ production and consumption primarily occurs in conjunction with methanolisation processes. Figures \ref{fig:PCI_lt_2030_h2}, \ref{fig:PCI_lt_2040_h2}, and \ref{fig:PCI_lt_2050_h2} provide a map of the regional distribution of H$_2$ production, utilisation, and transport in the \textit{PCI} scenario.

\paragraph{Pipeline delays beyond 2040 entail significant economic regret}
In this section, we discuss the impact of the three short-term scenarios on the long-term decarbonisation pathways, by comparing the economic regret, as well as the effects on CO$_2$ utilisation, sequestration, and H$_2$ production. We calculate the regret terms by subtracting the annual total system costs of the long-term scenarios (row) from the short-term scenarios (columns). In our analysis, regret values represent the additional costs incurred by a given short-term scenario relative to the benchmark. Positive values indicate higher costs, driven by increased investments in alternative generation, conversion, storage, and CDR technologies, as well as changes in their operation due to (i) delays or (ii) cancellations of pipeline infrastructure including PCI-PMI projects. Negative values indicate cost savings, which may arise under relaxed policy ambitions—for example, when CO$_2$ and H$_2$ targets are removed in the \textit{Reduced targets} scenario.

\begin{figure}[htbp]
  \centering
  \includegraphics{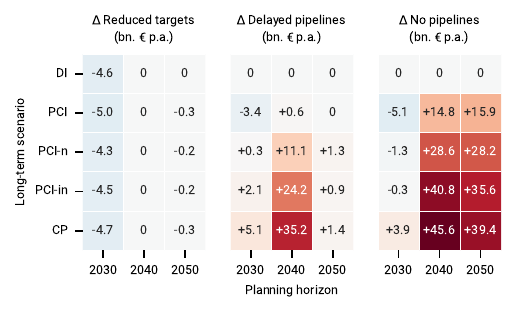}
  \caption{Regret matrix. \textit{Positive values indicate higher costs, driven by increased investments in alternative generation, conversion, storage, and CDR technologies, as well as changes in their operation due to (i) delays or (ii) cancellations of pipeline infrastructure including PCI-PMI projects. Negative values indicate cost savings, which may arise under relaxed policy ambitions—for example, when CO$_2$ and H$_2$ targets are removed in the \textit{Reduced targets} scenario.}}
  \label{fig:regret_matrix_results}
\end{figure}

Figure \ref{fig:regret_matrix_results} shows the regret for all scenarios and planning horizons. From left to right, the first column shows the regret terms for the \textit{Reduced targets} scenario, where all long-term targets are removed except for the GHG emission reduction target. The second column shows the regret terms for the \textit{Delayed pipelines} scenario, where all PCI-PMI and endogenous pipelines are delayed by one period. The third column shows the regret terms for the \textit{No pipelines} scenario, where all hydrogen and CO$_2$ pipeline capacities are removed.
In the \textit{Reduced targets} scenario, overall system costs change only marginally despite the relaxation of specific targets. This is because CO$_2$ sequestration levels are primarily driven by the overarching GHG emission constraints --- particularly the stringent 2040 and 2050 carbon budgets, which remain in place. With regard to hydrogen, the long-term results have previously shown that H$_2$ production targets were overachieved in 2040. Only in 2030, we see a net negative regret of around €4.3 to €4.6 billion per year, as the minimum H$_2$ production target was binding in the long-term scenario. Across all long-term scenarios, we have observed that CO$_2$ pipeline infrastructure is not essential in 2030 (see Figure \ref{fig:CP_lt_2030_co2}). In the case of H$_2$ pipeline infrastructure, the solution appears relatively flat: regrets in the \textit{DI} scenario without any pipelines (Figure \ref{fig:DI_lt_2030_co2}) are nearly identical to those in the \textit{CP} scenario (Figure \ref{fig:CP_lt_2030_co2}) with substantial pipeline deployment. When the H$_2$ production and CO$_2$ sequestration targets are removed, pipelines become even less relevant, although the associated cost savings are minimal, ranging from €4.3 to €5 billion per year in 2030 and 2040.

For similar reasons, the 2030 results for the \textit{Delayed pipelines} and \textit{No pipelines} scenarios exhibit small regret terms. Cost savings of €3.4 to €5.1 billion per year in the \textit{PCI} scenario suggest that, for 2030, mandating PCI-PMI projects is neither cost- nor topologically optimal in the short term. In contrast, a regret of €3.9 to €5.1 billion per year in the \textit{CP} scenario indicates some dependency on the invested pipeline infrastructure (Figure \ref{fig:CP_lt}) which represents the systemically more optimised solution.

When looking at the more long-term perspective, we see significant regrets in the \textit{Delayed pipelines} and \textit{No pipelines} scenarios. Having originally planned the energy system layout (including generation, transport, conversion technologies and storage) in the long-term scenario with PCI-PMI projects and/or endogenous pipelines, the model has to find alternative investments to still meet all targets, as the pipelines now materialise one period later or not at all. Regrets peak in 2040, where a delay of pipelines costs the system between €0.6 to €24.2 billion per year. in the scenarios with PCI-PMI projects and up to €35.2 billion p.a. in the \textit{CP} scenario. 2050 regrets are lower than 2040 regrets, as almost all PCI-PMI pipelines are originally commissioned by 2030. Hence, a delay of projects from 2040 to 2050 only mildly impacts the system costs by €0.6 billion per year. The more pipelines invested beyond those of PCI-PMI projects, the higher the regret if they are delayed. In 2050, very few additional CO$_2$ and H$_2$ pipelines are built, as such, a delay only increases system costs by €0.9 to €1.4 billion per year. 
The short-term scenario \textit{No pipelines} shows the highest regrets, ranging from €14.8 to €45.6 billion per year in 2040 and €15.9 to €39.4 billion per year in 2050. Note that this scenario represents a hypothetical worst case, as it is highly unlikely to plan an energy system with pipeline investments in mind yet fail to implement any of them.

Consistently throughout all short-term scenarios, most of the additional cost stem from the need to invest into additional carbon capture, renewable generation, and conversion technologies (see Figure \ref{fig:capacities_overview_extended}). Additional renewable generation capacities are made up of solar PV and wind. A significant higher amount of electrolyser capacity of more than 50 GW is needed in 2040 if pipelines are delayed. 

\paragraph{System pivots to high-cost DAC in the absence of pipeline infrastructure}
Further, the model has to invest in more than 28 GW of carbon capture units at point sources and an additional 14 GW in DAC technologies to meet the sequestration and emission reduction targets. Cost-wise, the short-term investments into DAC technologies make up to a half of the of the additional system costs in both the \textit{Delayed pipelines} and \textit{No pipelines} scenarios (see Figure \ref{fig:costs_overview_extended}). DAC utilisation can increase from 40 Mt p.a. in the \textit{PCI-n} to more than 200 Mt p.a. in the \textit{CP} scenario when pipelines are delayed (see Figure \ref{fig:balances_overview_extended_co2_stored}). If pipelines are not built at all, additional 60 Mt p.a. in the \textit{PCI} up to 250 Mt p.a. in the \textit{CP} scenario are captured from DAC, substituting a large share of CO$_2$ previously captured from point sources equipped with carbon capture (biomass-based industry processes and non-abatable process emissions).

\begin{figure}[htbp]
  \centering
  \includegraphics{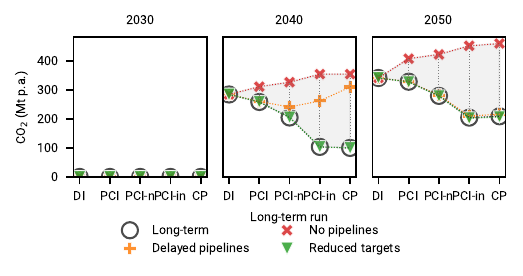}
  \caption{Delta balances --- CO$_2$ from DAC.}
  \label{fig:delta_balances_dac}
\end{figure}

Note that a clear trade-off between the reliance on pipeline infrastructure and the need for DAC technologies can be observed in Figure \ref{fig:delta_balances_dac}. While the reliance on DAC decreases with the build-out of pipeline infrastructure, the model in return has to invest in more DAC if pipelines are delayed or not built at all. There is a risk involved, that the need for DAC is even higher in the scenarios with pipeline infrastructure compared to the \textit{DI} scenario, especially in later years (2040 and 2050), if the pipelines do not materialise at all, seeing a potential increase of 50 Mt p.a. in 2040 and 80 Mt p.a. in 2050 in the \textit{PCI} scenario.

\paragraph{Hydrogen production shifts to more costly SMR and decentral production when pipelines are delayed} 
We find that the electrolytic H$_2$ production target of 10 Mt p.a. (333 TWh p.a.) in 2030 is overly ambitious. Figure \ref{fig:balances_overview_extended_H2_stored} shows that in the \textit{Reduced targets} scenario, 132 to 151 TWh p.a. of H$_2$, corresponding to almost half of the target is produced from SMR instead of electrolysis. When pipelines are delayed, the model has to fall back to more decentral H$_2$ production of an additional 55 to 187 TWh p.a. of H$_2$ from electrolysis, SMR and SMR with carbon capture (the latter being the most expensive option). In the \textit{No pipelines} scenario, this additional H$_2$ production increases to up to 305 TWh p.a (see Figure \ref{fig:balances_overview_extended_H2_stored}).

\begin{figure}[t]
  \centering
  \includegraphics{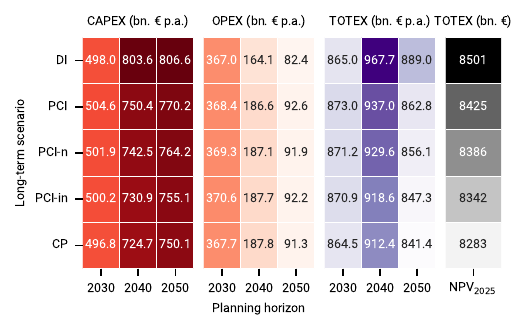}
  \caption{Annual system costs by long-term scenario and planning horizon.}
  \label{fig:totex_heatmap}
\end{figure}

\paragraph{In the long-run, PCI-PMI infrastructure deliver net system cost savings}
Looking at the long-run we find that PCI-PMI projects, while not completely cost-optimal compared to a centrally planned system, are still cost-beneficial. Compared to a complete lack of H$_2$ and CO$_2$ pipeline infrastructure as well as lower CO$_2$ sequestration potential, the \textit{PCI} scenario unlocks annual cost savings in up to €30.7 billion per year. Figure \ref{fig:totex_heatmap} shows the total system costs or Total Expenditures (TOTEX) p.a. split into Capital (CAPEX) and Operational Expenditures (OPEX) p.a., as well as the Net Present Value (NPV) of total system costs, discounted at an interest rate of 7\% p.a.
Even when accounting for the additional costs of €0.6 billion per year faced in the \textit{Delayed pipelines} and up to €15.9 billion per year in the \textit{No pipelines} scenario, a net positive is achieved, indicating that investing into the PCI-PMI infrastructure is a no-regret option. By connecting further H$_2$ production sites and CO$_2$ point sources to the pipeline network. additional cost savings of up to €18.4 billion per year can be achieved in the \textit{PCI-in} scenario. The \textit{CP} scenario serves as a theoretical benchmark, allowing the model to invest freely, not bound by \textit{forced} PCI-PMI projects. The model can invest in fewer, but more optimally located CO$_2$ and H$_2$ pipelines from a systemic perspective. Economic benefits of all pipeline investments materialise after 2030, yielding lower NPV of potentially at least €75 billion over the course of the assets' lifetime. 

\section*{Discussion}
\label{sec:discussion}

In this study, we have assessed the impact of PCI-PMI projects on reaching European climate targets on its path to net-zero by 2050. We have modelled the European energy system with a focus on H$_2$ and CO$_2$ infrastructure, and evaluated the performance of different levels of pipeline roll-out under three short-term scenarios. 

While our study assesses a variety of topologies, planning horizons, and potential regret scenarios, it is not exhaustive and comes with limitations. As we focus on the impact of continental European PCI-PMI infrastructure, we neglect fuel and energy imports from outside Europe. H$_2$ and CO$_2$ demand is directly driven by fixed, exogenous demands for the respective carrier or their derivatives. Regarding the modelling of both H$_2$ and CO$_2$ pipelines, we assume a level playing field for all pipeline projects through standardised costs and applying haversine distance, i.e., no discrimination between PCI-PMI projects and other projects, this is a simplification as real costs may differ. We also do not discretise the endogenously built pipelines (due to computational complexity) and allow any capacity to be built. This assumption can lead to underestimation of the true costs of pipeline investments.
Further, all results are based on a single weather year, i.e., 2013.
Other limitations include geographic and temporal clustering to make the problem solving computationally feasible.

\paragraph{Economic viability and policy targets}
Our findings demonstrate that PCI-PMI CO$_2$ and H$_2$ infrastructure generate a net positive impact on total system costs, even when accounting for potential additional costs involved with the delay of pipelines. This positions PCI-PMI projects as a no-regret investment option for the European energy system, when treated as a whole.
Their economic benefit increases considerably when strategic pipeline extensions are implemented, connecting additional H$_2$ production sites and CO$_2$ point sources to the pipeline network. 
Compared to a system without any pipeline infrastructure, PCI-PMI projects help to achieve the EU's ambitious policy targets, including net-zero emissions, H$_2$ production and CO$_2$ sequestration targets, while reducing system costs and technology dependencies.

\paragraph{CCUS and hydrogen utilisation}
The pipeline infrastructure serves dual purposes in Europe's decarbonisation strategy: H$_2$ pipelines facilitate the distribution of more affordable green H$_2$ from northern and south-western regions rich in renewable energy potential to high-demand regions in central Europe. Complementarily, CO$_2$ transport and offshore sequestration sites enable industrial decarbonisation by linking major industrial sites and their process emissions to offshore sequestration sites in the North Sea, particularly in Denmark, Norway, and the Netherlands.

\paragraph{Technology and risk diversification}
The build-out of CO$_2$ and H$_2$ pipeline infrastructure helps utilising renewable energy sources more efficiently. Hydrogen pipelines enable the transport of green H$_2$ over long distances while CO$_2$ pipelines reduce the reliance on single carbon capture technologies such as Direct Air Capture and point-source carbon capture, confirming the findings of \cite{hofmannH2CO2Network2025}. This diversification further enhances system resilience towards uncertainties involved with technologies that are not yet commercially available at scale, such as Direct Air Capture.

\paragraph{Political support and public acceptance} 
While PCI-PMI may not achieve perfect cost-optimality in their entirety compared to a theoretically centrally planned system, they possess benefits beyond pure economic viability. The success of large-scale infrastructure investments highly depend on continuous political support and public acceptance --- factors that are particularly favourable for PCI-PMI projects.
Backed directly by the European Commission, PCI-PMI projects benefit from stronger political endorsement, institutional support structures, enhanced access to financing and grants, and accelerated permitting processes. Additionally, the requirement for frequent and transparent progress reporting increases their likelihood of gaining public acceptance.

\section*{Methods}\label{sec:methods}
\paragraph{Overview of European energy system model PyPSA-Eur}
We build on the open-source, sector-coupled energy system model PyPSA-Eur \cite{neumannPotentialRoleHydrogen2023,frysztackiComparisonClusteringMethods2022,glaumOffshorePowerHydrogen2024,horschPyPSAEurOpenOptimisation2018} to optimise investment and dispatch decisions in the European energy system. The model's endogenous decisions include the expansion and dispatch of renewable energy sources, dispatchable power plants, power-to-X conversion, and storage/sequestration capacities as well as transmission infrastructure for power, hydrogen, and CO$_2$. It also encompasses heating technologies and various hydrogen production methods (gray, blue, green).
PyPSA-Eur integrates multiple energy carriers (e.g., electricity, heat, hydrogen, CO$_2$, methane, methanol, liquid hydrocarbons, and biomass) with corresponding conversion technologies across multiple sectors (i.e., electricity, transport, heating, biomass, industry, shipping, aviation, agriculture and fossil fuel feedstock). The model features high spatial and temporal resolution across Europe, incorporating existing power plant stocks \cite{gotzensPerformingEnergyModelling2019}, renewable potentials, and availability time series \cite{hofmannAtliteLightweightPython2021}. It includes the current high-voltage transmission grid (AC 220 to 750 kV and DC 150 kV and above) \cite{xiongModellingHighvoltageGrid2025}. Furthermore, electricity transmission projects from the TYNDP \cite{entso-eTenYearNetworkDevelopment2020} and German Netzentwicklungsplan \cite{bnetzaBestaetigungNetzentwicklungsplanStrom2024} are also enabled.

\paragraph{Geographical and temporal resolution}
To assess the long-term impact of PCI-PMI projects on European policy targets across all sectors, we optimise the sector-coupled network for three key planning horizons 2030, 2040, and 2050, myopically. The myopic approach ensures that investment decisions across all planning horizons are non-anticipative and build on top of the previous planning horizon. We use a time series aggregation technique to solve the model with 2190 representative time steps. The aggregation is done with the Python package \textit{tsam} developed by Kotzur et al. \cite{kotzurImpactDifferentTime2018} which ensures that intertemporal characteristics including renewable infeed variability, demand fluctuations, and seasonal storage needs are preserved.

We model 34 European countries, including 25 of the EU27 member states (excluding Cyprus and Malta), as well as Norway, Switzerland, the United Kingdom, Albania, Bosnia and Herzegovina, Montenegro, North Macedonia, Serbia, and Kosovo. Regional clustering is based on administrative NUTS boundaries, with higher spatial resolution applied to regions hosting planned PCI-PMI infrastructure, producing 99 onshore regions (see Table \ref{tab:regional_clustering}). Depending on the scenario, additional offshore buses are introduced to appropriately represent offshore sequestration sites and PCI-PMI projects. To isolate the effect of PCI-PMI projects, Europe is self-sufficient in our study, i.e., we do not allow any imports or exports of the assessed carriers like electricity, H$_2$, or CO$_2$. 

\paragraph{Technology assumptions}
As part of the PyPSA-Eur model, all technology-specific assumptions --- such as lifetime, efficiency, investment costs, and operational costs --- are derived from the public Energy System Technology Data repository (v0.10.1) \cite{zeyenPyPSATechnologydataV01012025}. This repository sources most of its data from technology catalogues published by the Danish Energy Agency (Energistyrelsen) \cite{energistyrelsendanishenergyagencyTechnologyCatalogues2024}.We use values projected for 2030 and apply a discount rate of 7\%, reflecting the weighted average cost of capital (WACC). We assume CO$_2$ sequestration costs of €15/tCO$_2$ which can be considered in the mid-range of the cost spectrum (cf. €10/tCO$_2$ \cite{hofmannH2CO2Network2025} and \$12 to \$18/tCO$_2$ \cite{rubinCostCO2Capture2015}). An overview of selected technology assumptions is provided in Table \ref{tab:cost_assumptions}.

\paragraph{Energy demand and CO$_2$ emissions}
Energy and fuel carrier demand in the modelled sectors, as well as non-abatable CO$_2$ process emissions are taken from various sources \cite{mantzosJRCIDEES20152018,eurostatCompleteEnergyBalances2022,manzGeoreferencedIndustrialSites2018,muehlenpfordtTimeSeries2019,krienOemofDemandlibV0222025} and are shown in Figure \ref{fig:exogenous_demand}. Regionally and temporally resolved demand includes electricity, heat, gas, biomass and transport. 

Gas (methane/CH$_4$) demand includes direct use in gas-based industrial processes, as well as fuel in the electricity and heating sector. Note that we do not explicitly model the gas transmission grid as opposed to the CO$_2$ and H$_2$ infrastructure. We do this for the following reasons: (i) The modelled PCI-PMI projects overlap in some parts with the gas grid, i.e., they include CH$_4$ pipelines that will be retrofitted to H$_2$ pipelines --- information in the PCI-PMI project sheets is not always clear on this; (ii) In the EU energy system, the transport of natural gas is rarely constrained by the existing gas grid infrastructure, reflecting the grid's robust capacity to accommodate demand fluctuations \cite{riepinModellingUncertaintyCoupled2021}; (iii) Considering (ii), empirical gains of explicitly implementing the gas grid do not justify the additional computational burden. 
Instead, given this work's focus on the CO$_2$ and H$_2$ sector, we have decided to make trade-offs here and assume gas transport to be `copper-plated'.

Internal combustion engine vehicles in land transport are expected to fully phase out in favour of electric vehicles by 2050 \cite{zeyenShiftingBurdensHow2025a}. Demand for methanol and hydrocarbons, e.g. kerosene, are primarily driven by the shipping, aviation and industry sector and are not spatially resolved (Figure \ref{fig:exogenous_demand}).
To reach net-zero CO$_2$ emissions by 2050, the yearly emission budget follows the EU's 2030 (-55\%) and 2040 (-90\%) targets \cite{europeancommissionFit55Delivering2021, europeancommission.directorategeneralforclimateaction.IndepthReportResults2024}, translating into a carbon budget of 2072 Mt p.a. in 2030 and 460 Mt p.a. in 2040, respectively (see Table \ref{tab:targets}).

\paragraph{PCI-PMI H$_2$ and CO$_2$ transport, storage, and sequestration}

We implement all PCI-PMI projects of the electricity, CO$_2$ and H$_2$ sectors (excluding offshore energy islands and hybrid interconnectors, as they are not the focus of our research) by accessing the REST API of the PCI-PMI Transparency Platform and associated public project sheets provided by the European Commission \cite{europeancommissionPCIPMITransparencyPlatform2024}. We add all CO$_2$ sequestration sites and connected pipelines, H$_2$ pipelines and storage sites, as well as proposed pumped-hydro storage units and transmission lines (AC and DC) to the PyPSA-Eur model (see Figure \ref{fig:regional_scope_map}). We consider the exact geographic information, build year, as well as available static technical parameters when adding individual assets to the respective modelling year. 

Our implementation can adapt to the needs and configuration of the model, including selected technologies, geographical and temporal resolution, as well as considered sectors. Within this study, all projects are mapped to the 99 NUTS regions. In the mapping process, pipelines are aggregated and connect all regions that they are overpassing. Similar to how all electricity lines and carrier links are modelled in PyPSA-Eur, lengths are calculated using the haversine formula multiplied by a factor of 1.25 to account for the non-straight shape of pipelines.
We apply standardised cost assumptions \cite{zeyenPyPSATechnologydataV01012025} across all existing brownfield assets, exogenously specified PCI-PMI projects, and projects endogenously selected by the model, equally. Our approach is motivated by two considerations: (i) cost data submitted by project promoters are often incomplete and may differ in terms of included components, underlying assumptions, and risk margins; and (ii) applying uniform cost assumptions ensures comparability and a level playing field across all potential investments, including both PCI-PMI projects and endogenous model decisions.

Beyond CO$_2$ sequestration site projects included in the latest PCI-PMI list (around 114 Mt p.a.), we consider additional technical potential from the European CO$_2$ storage database \cite{europeancommissionEuropeanCO2Storage2020,hofmannH2CO2Network2025}. The dataset includes storage potential from depleted oil and gas fields and saline aquifers. While social and commercial acceptance of CO$_2$ storage has been increasing in recent years, concerns still exist regarding its long-term role and safety \cite{vanalphenSocietalAcceptanceCarbon2007}.
We only consider conservative estimates from depleted oil and gas fields, which are primarily located offshore in the British, Norwegian, and Dutch North Sea (see Figure \ref{fig:regional_scope_map}), yielding a technical sequestration potential of 7164 Mt. Our focus is motivated by the following reasons: (i) infrastructure such as wells, platforms, and pipelines already exist for depleted oil and gas fields and can be repurposed, significantly lowering costs and project risk; (ii) depleted fields are generally better understood geologically and have demonstrated sealing capacities, further reducing uncertainty; and (iii) repurposing former production sites is often more publicly and politically acceptable than developing entirely new storage locations, entirely. In contrast, while saline aquifers represent a substantial share of the total technical potential, they carry higher development costs and risks and are less likely to be advanced without strong policy and financial support \cite{europeancommissionEuropeanCO2Storage2020}. Note that the PCI-PMI project list includes some aquifer-based sequestration projects, however, their inclusion as PCI-PMI project indicates a higher likelihood of development.
We distribute the technical sequestration potential of the depleted oil and gas fields over a lifetime of 25 years (cf. \cite{hofmannH2CO2Network2025}), yielding an annual sequestration potential of up to 286 Mt p.a. We then cluster all offshore potential within a buffer radius of 50 km per offshore bus region in each modelled NUTS region and connect them through offshore CO$_2$ pipelines to the closest onshore bus. 
The model also includes H$_2$ storage sites from the PCI-PMI list and allows for endogenous build-out of additional storage capacities by repurposing salt caverns \cite{neumannPotentialRoleHydrogen2023}.

\paragraph{Climate and energy policy targets}
In all long-term scenarios, emission, technology, sequestration and production targets have to be met for each planning horizon (see Table \ref{tab:targets}). For the year 2030, these targets are directly derived from the EU's policy targets, including a 55\% reduction in greenhouse gas emissions compared to 1990 levels \cite{europeancommissionFit55Delivering2021}, 10 Mt p.a. of domestic green H$_2$ production \cite{europeancommissionREPowerEUPlanCommunication2022} and 40 GW of electrolyser capacity \cite{europeancommissionCommunicationCommissionEuropean2020}, and 50 Mt p.a. of CO$_2$ sequestration capacity \cite{europeanparliamentRegulationEU20242024}. For 2050, the CO$_2$ are based on the modelling the impact assessment for the EU's 2040 climate targets, in 250 Mt p.a. need to be sequestered \cite{europeancommissionCommunicationCommissionEuropean2024}. H$_2$ production targets for 2050 are based on the European Commission's \textit{METIS 3 study S5} \cite{europeancommission.directorategeneralforenergy.METIS3Study2023}, modelling possible pathways for industry decarbonisation until 2040. For 2040, we interpolate linearly between the 2030 and 2050 targets. The electrolyser capacities for 2040 and 2050 are scaled by the ratio of H$_2$ production to electrolyser capacity in 2030. An overview of the targets and their values is provided in Table \ref{tab:targets}. We implement the green hydrogen production target as a minimum production constraint on electrolysis. Accordingly, we refer to this hydrogen as \textit{electrolytic H$_2$} throughout this paper. Note that this implementation is based on an aggregated annual target without temporal matching rules.

\begin{table}[htbp]
  \centering
  \caption{Pathway for implemented targets. Climate and energy policy targets based on \cite{europeancommissionFit55Delivering2021,europeancommissionREPowerEUPlanCommunication2022,europeanparliamentRegulationEU20242024,europeancommissionCommunicationCommissionEuropean2024,europeancommission.directorategeneralforenergy.METIS3Study2023}}
  \label{tab:targets}
  \scriptsize
  \begin{tabularx}{\linewidth}{R{4.2cm}>{\centering\arraybackslash}X>{\centering\arraybackslash}X>{\centering\arraybackslash}X}
    \toprule
    \textbf{Planning horizon} & \textbf{2030} & \textbf{2040} & \textbf{2050} \\
    \midrule
    \textbf{Targets} & & & \\
    GHG emission reduction &  -55\% & -90\% & -100\% \\
    CO$_2$ sequestration & 50 Mt p.a. & 150 Mt p.a. & 250 Mt p.a. \\
    Electrolytic H$_2$ production & 10 Mt p.a. & 27.5 Mt p.a. & 45 Mt p.a. \\
    H$_2$ electrolyser capacity & 40 GW &  110 GW &  180 GW \\
    \bottomrule
  \end{tabularx}
\end{table}

\backmatter

\section*{Data availability}
A dataset of the model results is published on Zenodo under \href{https://doi.org/10.5281/zenodo.15790593}{https://doi.org/10.5281/zenodo.15790593} \cite{xiongResultDataRole2025}. Data on techno-economic assumptions can be found at \href{https://github.com/PyPSA/technology-data}{https://github.com/PyPSA/technology-data}

\section*{Code availability}
The entire code and workflow to reproduce the results is available at \href{https://github.com/bobbyxng/pci-pmi-policy-targets}{https://github.com/bobbyxng/pci-pmi-policy-targets}, which is based on a derivative of PyPSA-Eur v2025.01.0. We also refer to the documentation of PyPSA-Eur at \href{https://pypsa-eur.readthedocs.io/en/v2025.01.0}{https://pypsa-eur.readthedocs.io/en/v2025.01.0} for more information on the model.

\section*{Acknowledgements}
B.X. and I.R. gratefully acknowledge funding from the RESILIENT project supported by the German Federal Ministry for Economic Affairs and Climate Action (BMWK) under Grant No. 03EI4083A. 

\section*{Author contributions}
B.X.: Conceptualisation, Methodology, Software, Validation, Investigation, Data Curation, Writing --- Original Draft, Review \& Editing, Visualisation. I.R.: Conceptualisation, Methodology, Investigation, Writing --- Review \& Editing, Project Administration, Supervision. T.B.: Investigation, Resources, Writing --- Review \& Editing, Supervision, Funding acquisition.

\section*{Funding}
This project has been funded by partners of the CETPartnership (\href{https://cetpartnership.eu}{https://cetpartnership.eu}) through the Joint Call 2022. As such, this project has received funding from the European Union's Horizon Europe research and innovation programme under grant agreement no. 101069750.

\section*{Competing interests}
The authors declare no competing interests.

\clearpage
\begin{appendices}

\section{Data \& methodology}\label{app:data_methodology}

\begin{figure}[htbp]
  \centering
  \includegraphics{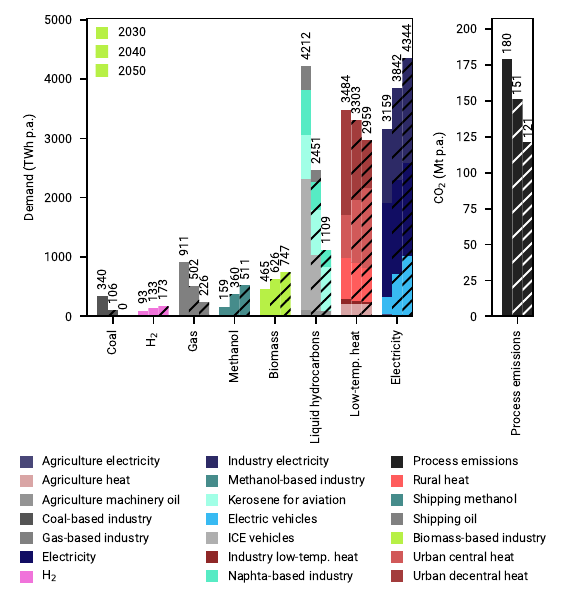}
  \caption{Exogenous demand.}
  \label{fig:exogenous_demand}
\end{figure}

\begin{table}[htbp]
  \centering
  \caption{Cost assumptions for key technologies based on \cite{zeyenPyPSATechnologydataV01012025}.}
  \label{tab:cost_assumptions}
  \scriptsize
  \begin{tabularx}{\linewidth}{R{4.2cm}>{\centering\arraybackslash}X>{\centering\arraybackslash}X>
  {\centering\arraybackslash}X}
    \toprule
    & \textbf{Unit} & \textbf{CAPEX} & \textbf{FOM} \\
    \midrule
    \textbf{Pipeline infrastructure} & & & \\
    CO$_2$ onshore pipelines & €/tCO$_2$/hkm & 2116 & 0.9\%/a \\
    CO$_2$ offshore pipelines & €/tCO$_2$/hkm & 4233 & 0.5\%/a \\
    H$_2$ onshore pipelines & €/MW$_{H_2}$/km & 304 & 1.5-3.2\%/a \\
    H$_2$ offshore pipelines & €/MW$_{H_2}$/km & 456 & 3.0\%/a \\
    \midrule
    \textbf{Conversion} & & & \\
    Electrolysis & €/kW$_e$ & 1000-1500 & 4.0\%/a \\
    SMR & €/MW$_{CH_4}$ & 522201 & 5.0\%/a \\
    SMR CC & €/MW$_{CH_4}$ & 605753 & 5.0\%/a \\

    \bottomrule
  \end{tabularx}
\end{table}

\begin{table}[htbp]
  \centering
  \caption{Regional clustering: A total of 99 regions are modelled, excluding offshore buses.}
  \label{tab:regional_clustering}
  \scriptsize
  \begin{tabularx}{\linewidth}{R{3cm}>{\centering\arraybackslash}X>{\centering\arraybackslash}X}
    \toprule
     & \textbf{Country} & \textbf{Buses} \\
    \midrule
    \textbf{Admin. level} & $\bm\sum$ & \textbf{99} \\
    NUTS2 & Finland (FI) & 4 \\
          & Norway (NO) & 6 \\
    \midrule
    NUTS1 & Belgium (BE)\footnotemark[1] & 2 \\
          & Switzerland (CH) & 1 \\
          & Czech Republic (CZ) & 1 \\
          & Germany (DE)\footnotemark[1] & 13 \\
          & Denmark (DK) & 1 \\
          & Estonia (EE) & 1 \\
          & Spain (ES)\footnotemark[1] & 5 \\
          & France (FR) & 13 \\
          & Great Britain (GB)\footnotemark[1] & 11 \\
          & Greece (GR) & 3 \\
          & Ireland (IE) & 1 \\
          & Italy (IT)\footnotemark[1] & 6 \\
          & Lithuania (LT) & 1 \\
          & Luxembourg (LU) & 1 \\
          & Latvia (LV) & 1 \\
          & Montenegro (ME) & 1 \\
          & Macedonia (MK) & 1 \\
          & Netherlands (NL) & 4 \\
          & Poland (PL) & 7 \\
          & Portugal (PT) & 1 \\
          & Sweden (SE) & 3 \\
          & Slovenia (SI) & 1 \\
          & Slovakia (SK) & 1 \\
    \midrule
    NUTS0 & Albania (AL) & 1 \\
          & Austria (AT) & 1 \\
          & Bosnia and Herzegovina (BA) & 1 \\
          & Bulgaria (BG) & 1 \\
          & Croatia (HR) & 1 \\
          & Hungary (HU) & 1 \\
          & Romania (RO) & 1 \\
          & Serbia (RS) & 1 \\
          & Kosovo (XK) & 1 \\
    \bottomrule
  \end{tabularx}
  \centering
  \footnotetext{Sardinia and Sicily are modelled as two separate regions.}
  \footnotetext[1]{City-states, i.e., Berlin, Bremen, Hamburg, Madrid, London, and regions without substations are merged with neighbours.}
\end{table}

\clearpage
\section{Results}\label{app:results}

\begin{figure}[htbp]
  \centering
  \includegraphics{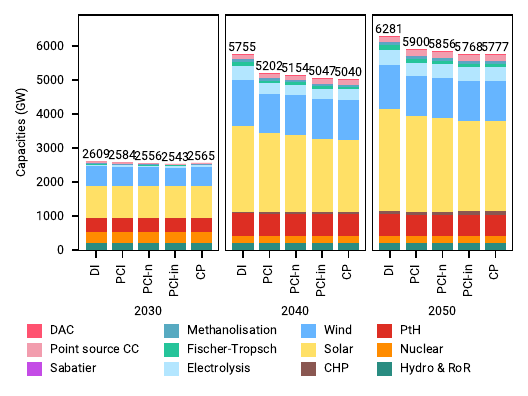}
  \caption{Installed capacities in long-term scenarios.}
  \label{fig:capacities_overview}
\end{figure}

\begin{figure}[htbp]
  \centering
  \includegraphics{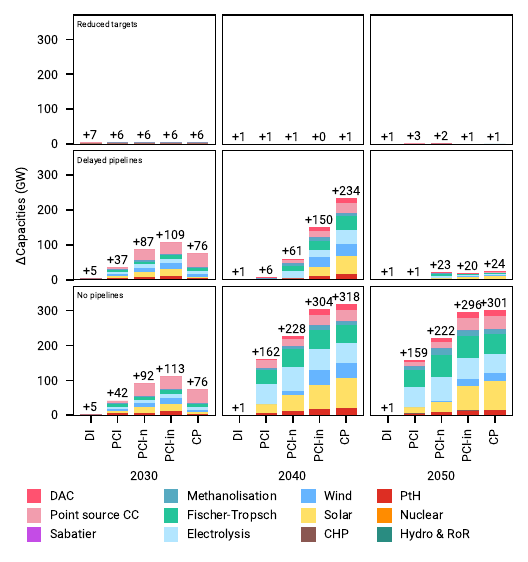}
  \caption{$\Delta$Capacities --- Short-term minus long-term runs.}
  \label{fig:capacities_overview_extended}
\end{figure}

\begin{figure}[htbp]
  \centering
  \includegraphics{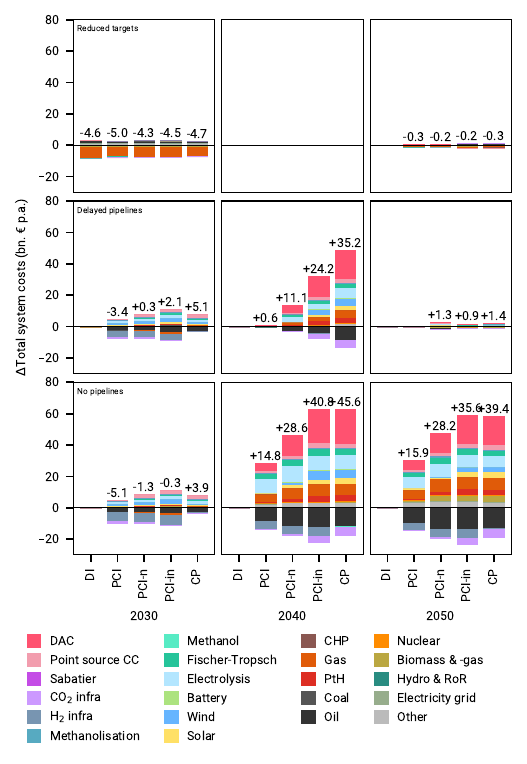}
  \caption{$\Delta$System costs --- Short-term minus long-term runs.}
  \label{fig:costs_overview_extended}
\end{figure}

\begin{figure}[htbp]
  \centering
  \includegraphics{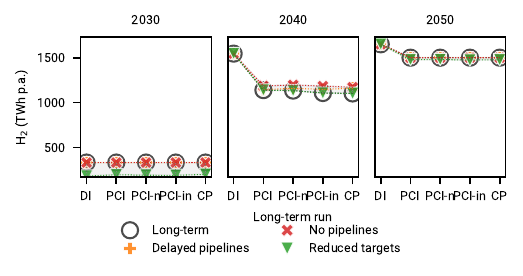}
  \caption{Delta balances --- Electrolytic H$_2$ production}
  \label{fig:delta_balances_h2_electrolysis}
\end{figure}

\begin{figure}[htbp]
  \centering
  \includegraphics{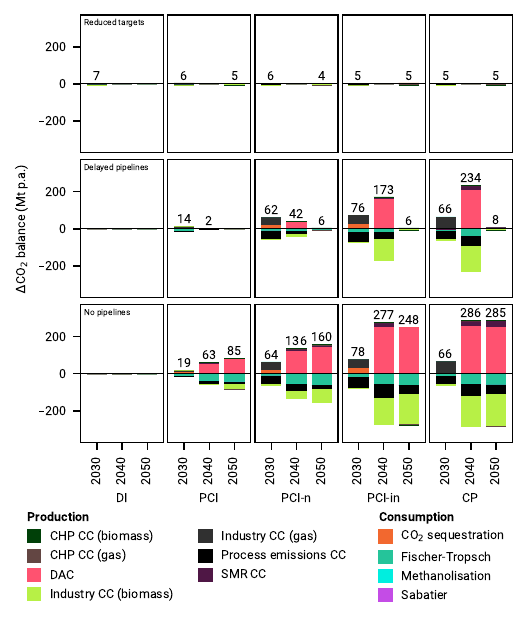}
  \caption{$\Delta$CO$_2$ balances --- Short-term minus long-term runs.}
  \label{fig:balances_overview_extended_co2_stored}
\end{figure}

\begin{figure}[htbp]
  \centering
  \includegraphics{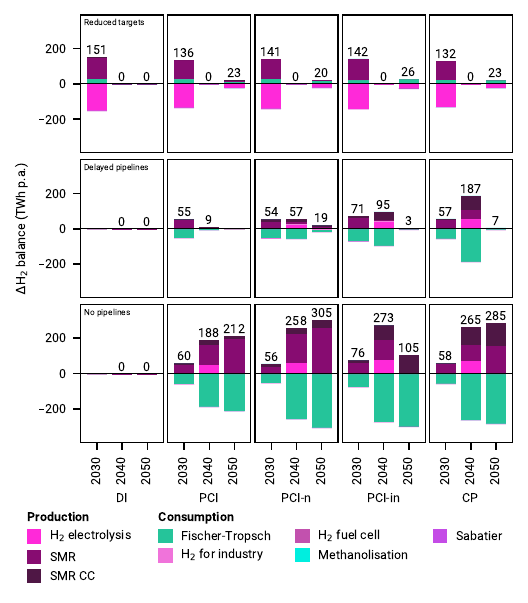}
  \caption{$\Delta$H$_2$ balances --- Short-term minus long-term runs.}
  \label{fig:balances_overview_extended_H2_stored}
\end{figure}

\begin{figure}[htbp]
  \centering
  \includegraphics{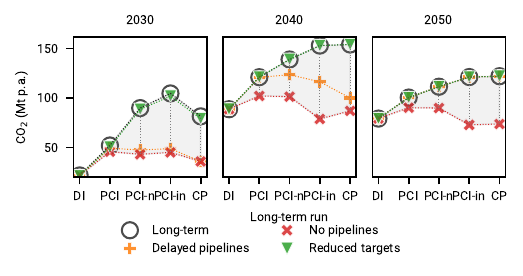}
  \caption{$\Delta$CO$_2$ balances --- Process emissions including carbon capture.}
  \label{fig:delta_balances_process_emissions_CC}
\end{figure}

\begin{figure}[htbp]
  \centering
  \includegraphics{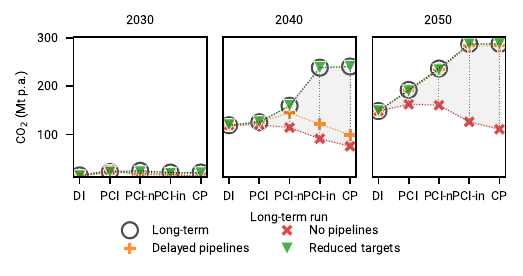}
  \caption{$\Delta$CO$_2$ balances --- Carbon capture from solid biomass for industry point sources.}
  \label{fig:delta_balances_biomass_industry_cc}
\end{figure}

\begin{figure}[htbp]
  \centering
  \includegraphics{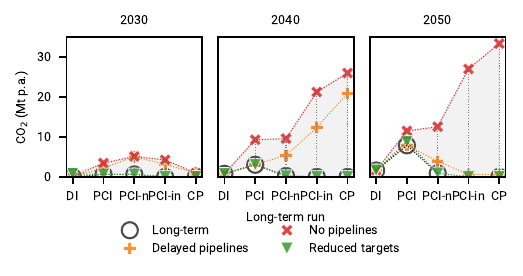}
  \caption{$\Delta$CO$_2$ balances --- Carbon capture from SMR point sources.}
  \label{fig:delta_balances_smr_cc}
\end{figure}

\begin{figure}[htbp]
  \centering
  \includegraphics{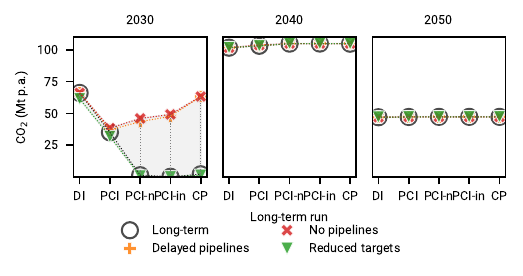}
  \caption{$\Delta$CO$_2$ balances --- Carbon capture from gas for industry point sources.}
  \label{fig:delta_balances_gas_for_industry}
\end{figure}

\begin{figure}[htbp]
  \centering
  \includegraphics{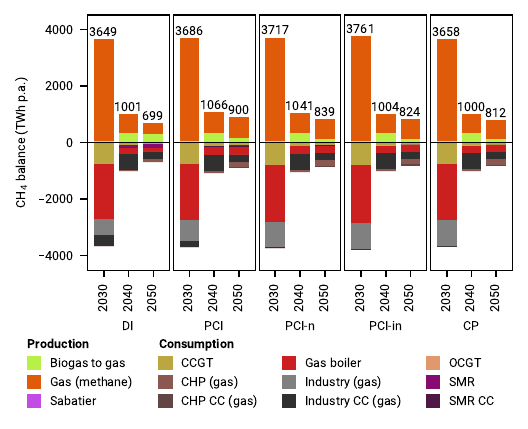}
  \caption{CH$_4$ balances in long-term scenarios.}
  \label{fig:balances_overview_gas}
\end{figure}

\begin{figure}[htbp]
  \centering
  \includegraphics{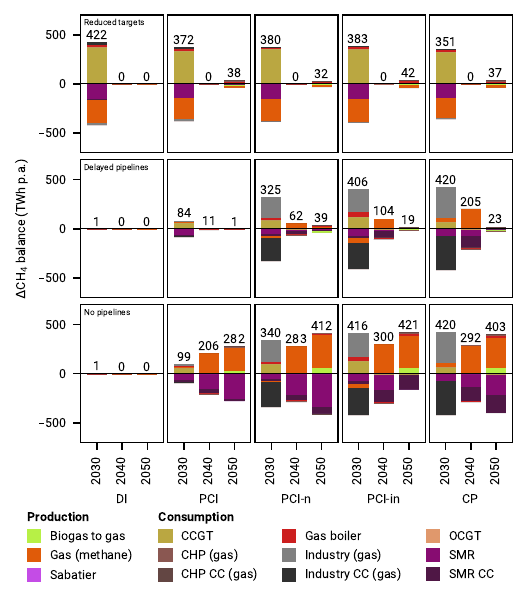}
  \caption{$\Delta$CH$_4$ balances --- Short-term minus long-term runs.}
  \label{fig:balances_overview_extended_gas}
\end{figure}

\begin{figure*}[htbp]
  \centering
  \begin{subfigure}[t]{0.32\textwidth}
      \vspace{0pt}
      \includegraphics[width=1\textwidth]{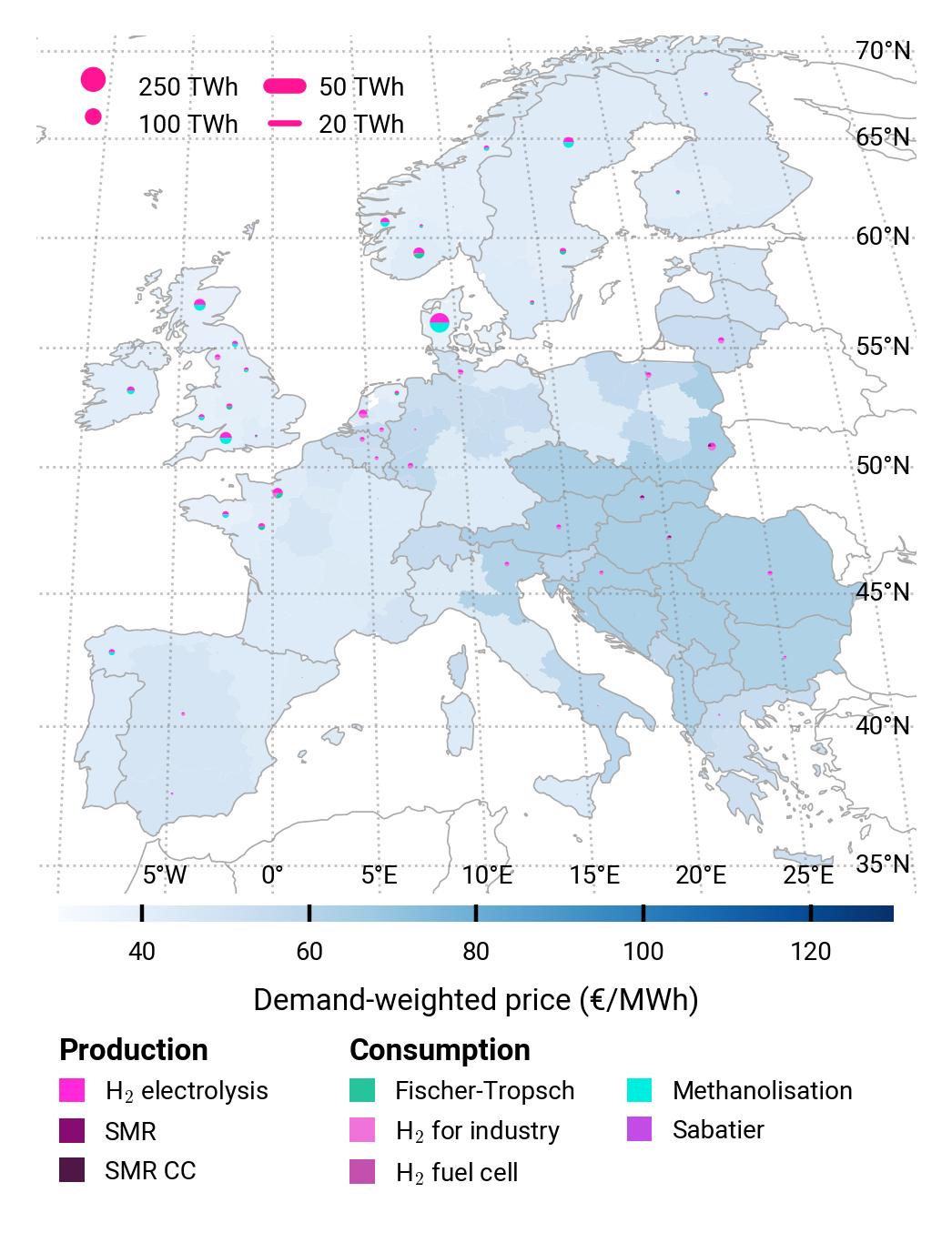}
      \caption{H$_2$ 2030.}
      \label{fig:DI_lt_2030_h2}
  \end{subfigure}
  \begin{subfigure}[t]{0.32\textwidth}
      \vspace{0pt}
      \includegraphics[width=1\textwidth]{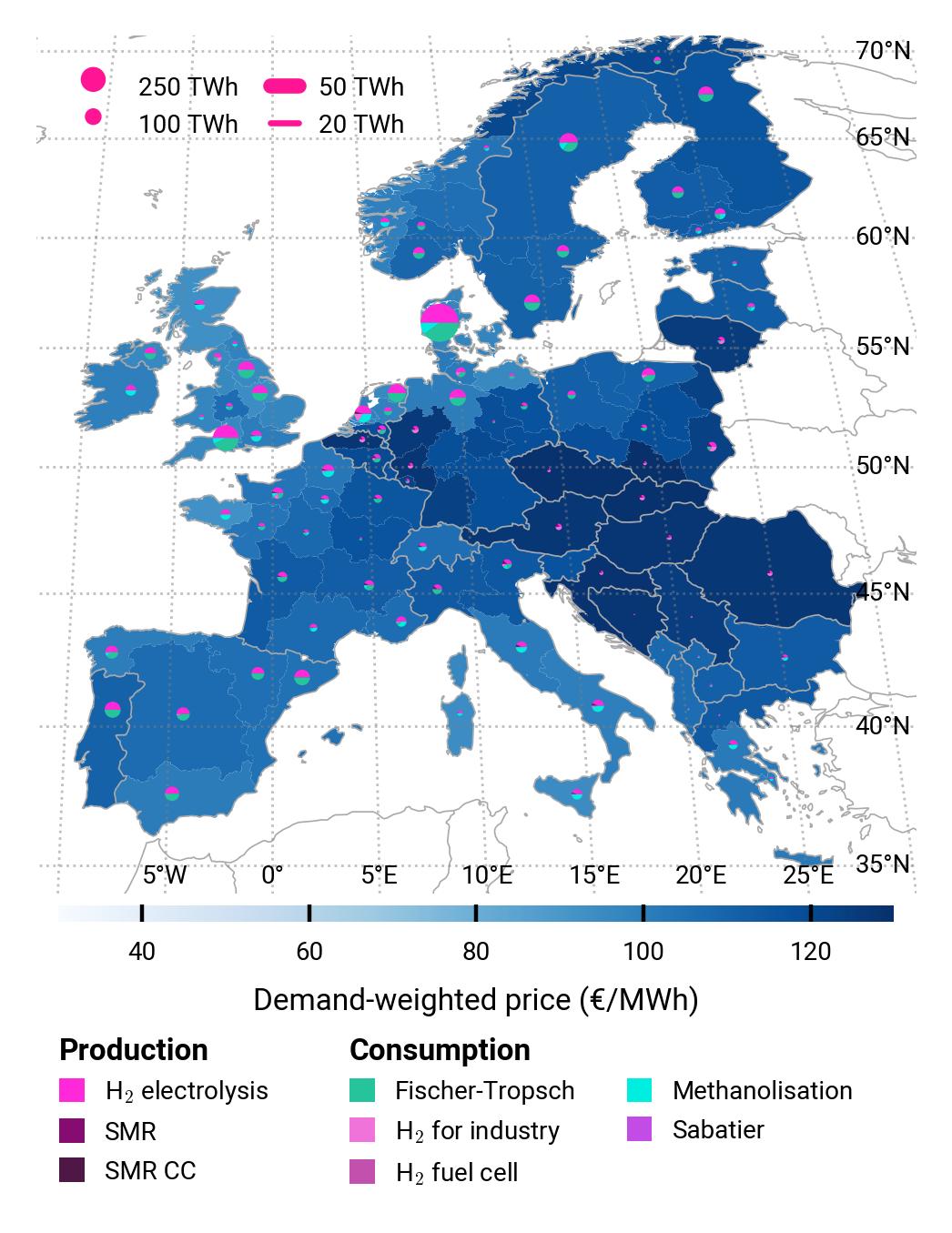}
      \caption{H$_2$ 2040.}
      \label{fig:DI_lt_2040_h2}
  \end{subfigure}
  \begin{subfigure}[t]{0.32\textwidth}
    \vspace{0pt}
    \includegraphics[width=1\textwidth]{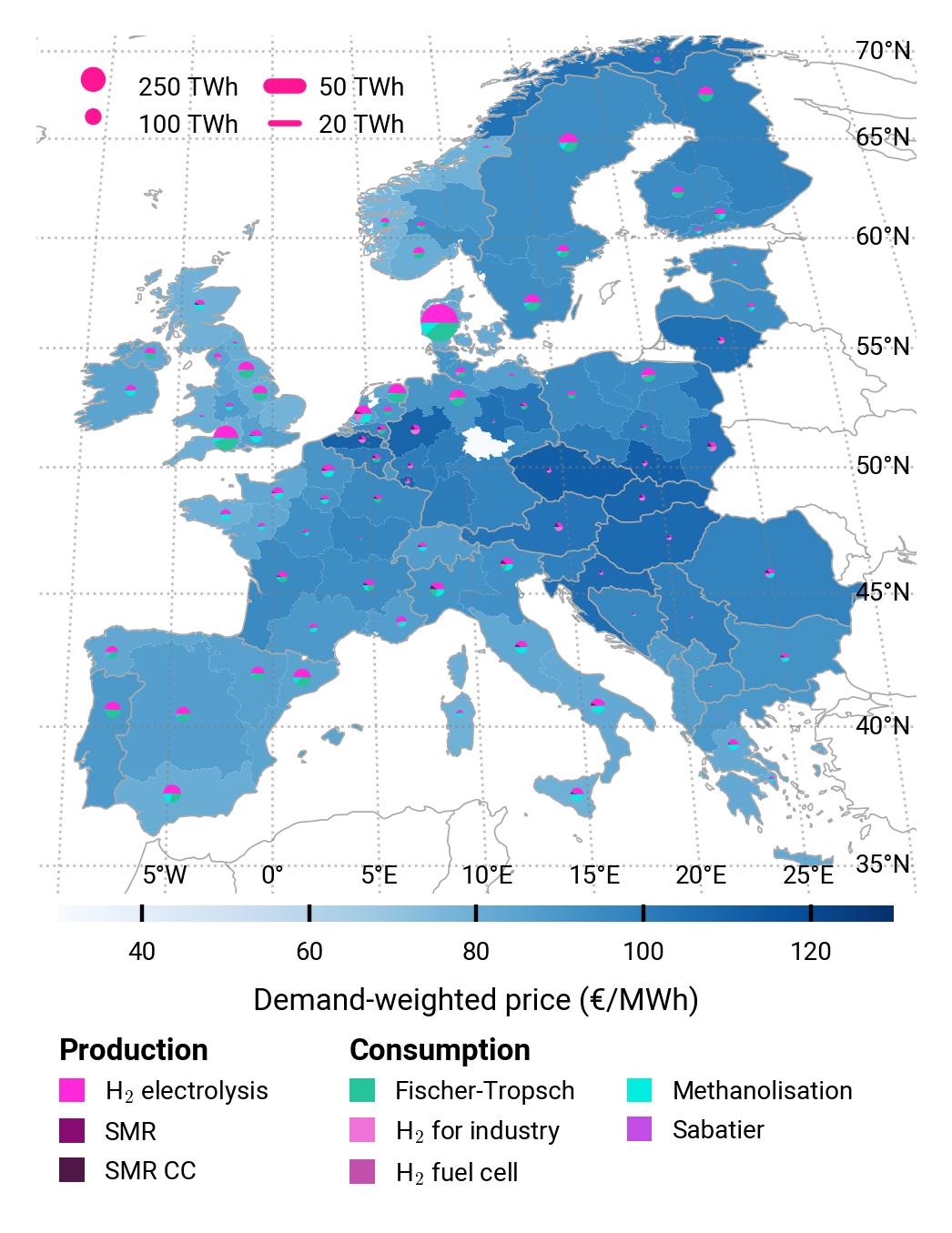}
    \caption{H$_2$ 2050.}
    \label{fig:DI_lt_2050_h2}
  \end{subfigure}
  \begin{subfigure}[t]{0.32\textwidth}
      \vspace{0pt}
      \includegraphics[width=1\textwidth]{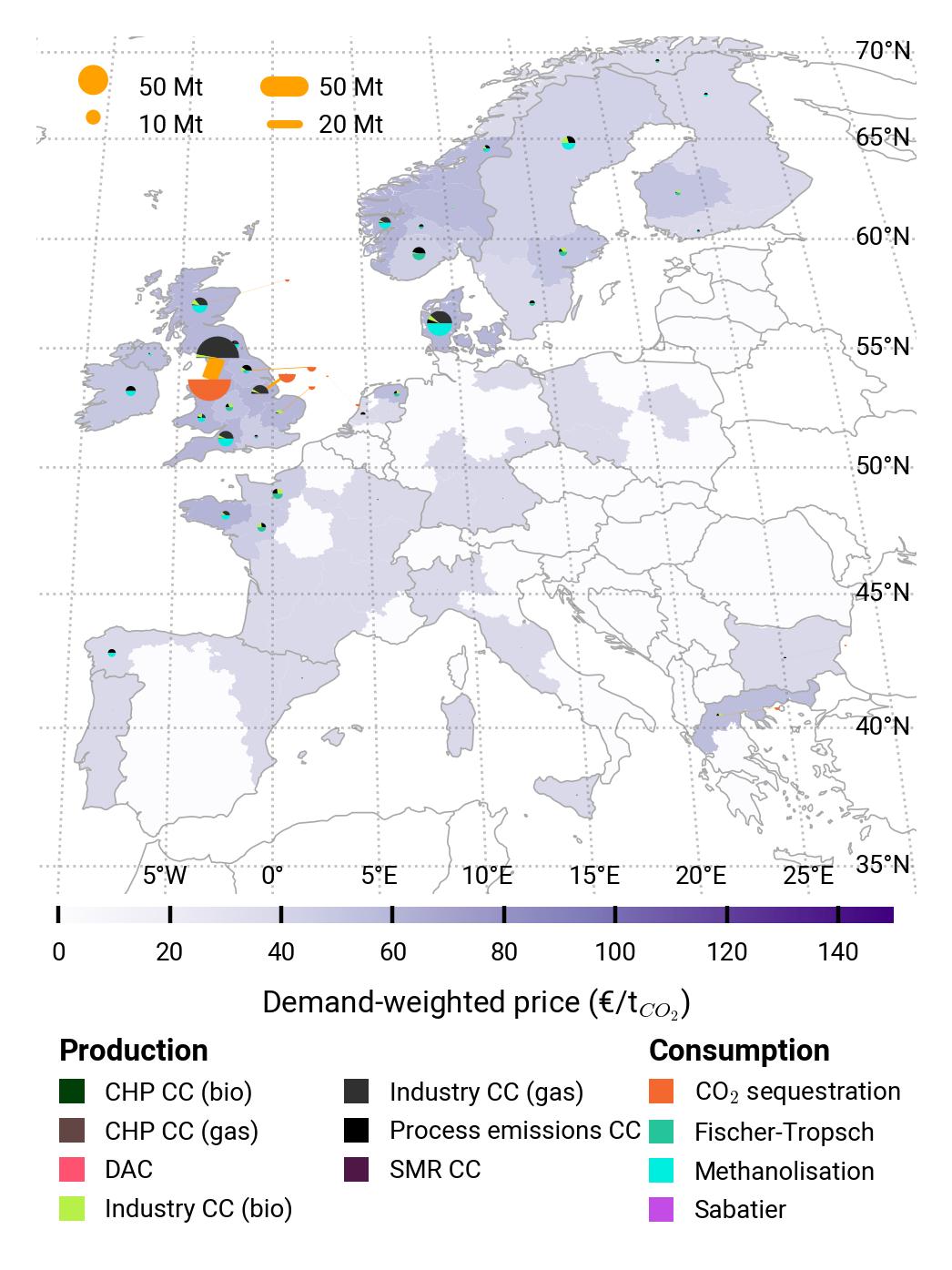} 
      \caption{CO$_2$ 2030.}
      \label{fig:DI_lt_2030_co2}
  \end{subfigure}
  \begin{subfigure}[t]{0.32\textwidth}
      \vspace{0pt}
      \includegraphics[width=1\textwidth]{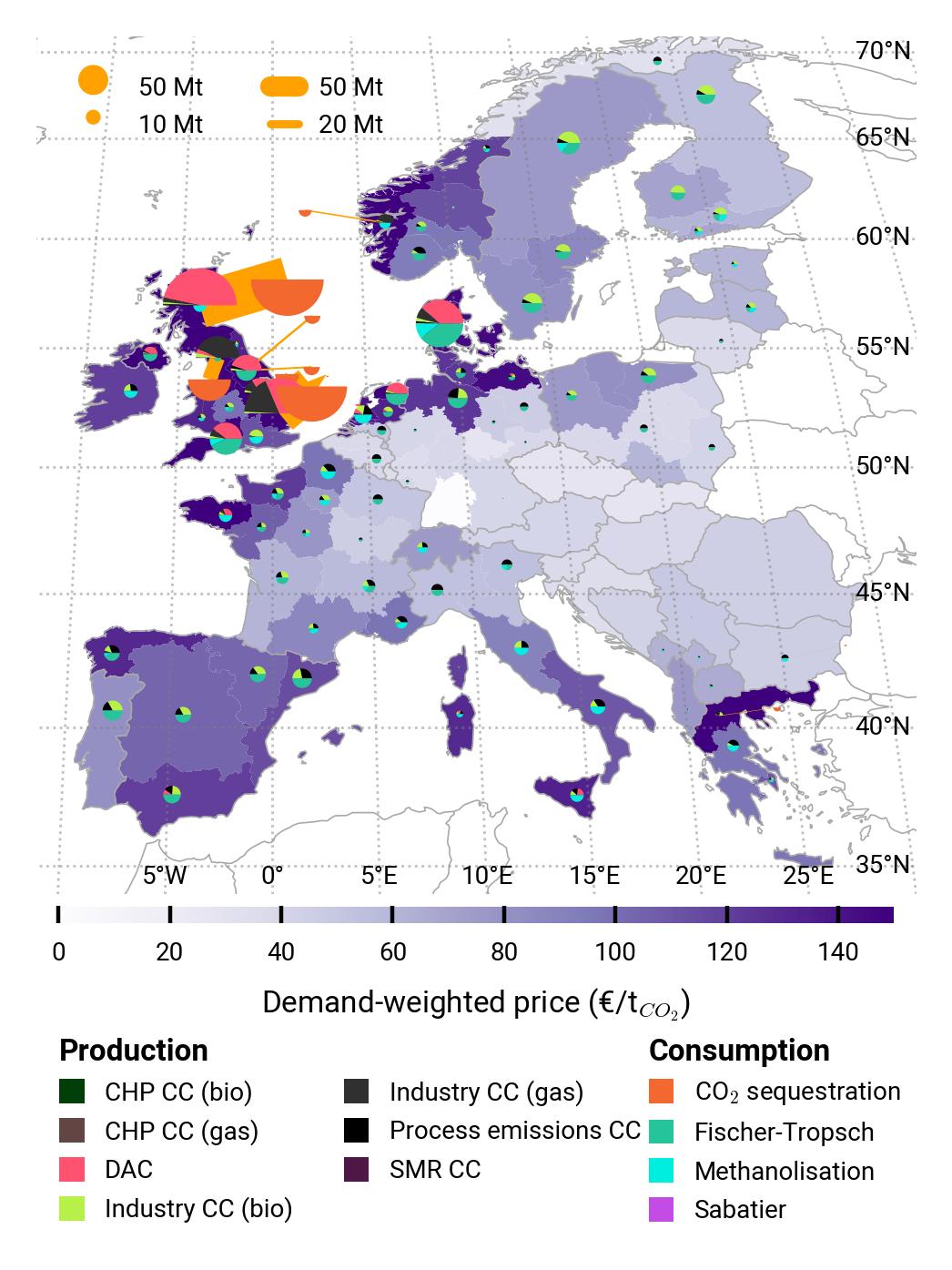} 
      \caption{CO$_2$ 2040.}
      \label{fig:DI_lt_2040_co2}
  \end{subfigure}
  \begin{subfigure}[t]{0.32\textwidth}
      \vspace{0pt}
      \includegraphics[width=1\textwidth]{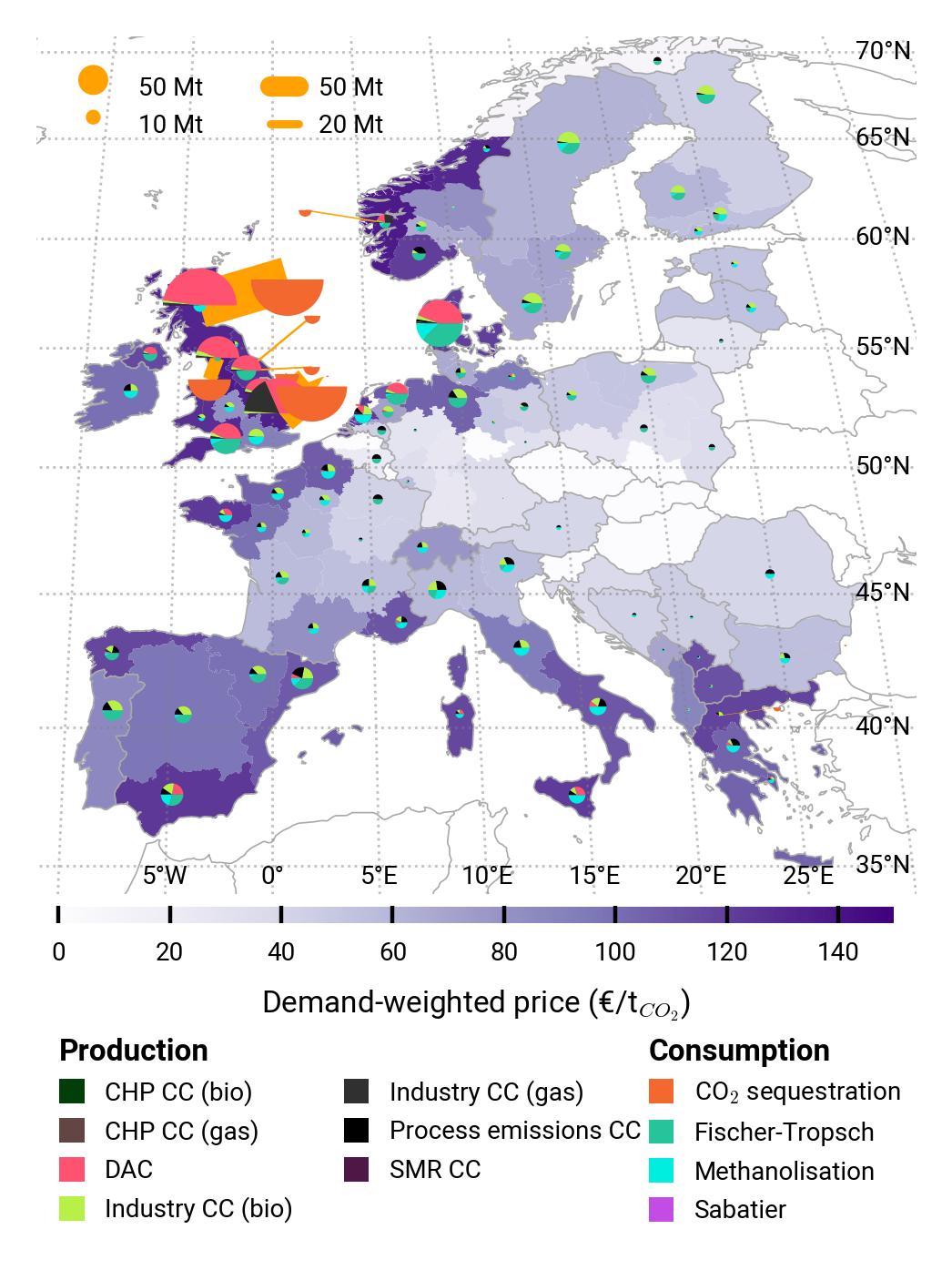} 
      \caption{CO$_2$ 2050.}
      \label{fig:DI_lt_2050_co2}
  \end{subfigure}
  \vspace{0.3cm}
  \caption{\textit{Decentral Islands} long-term scenario --- Regional distribution of H$_2$ and CO$_2$ production, utilisation, storage, transport and price. Note that both the H$_2$ and CO$_2$ price refer to their value as a commodity, i.e., price is higher where there is a demand for it.}
  \label{fig:DI_lt}
\end{figure*}

\begin{figure*}[htbp]
  \centering
  \begin{subfigure}[t]{0.32\textwidth}
      \vspace{0pt}
      \includegraphics[width=1\textwidth]{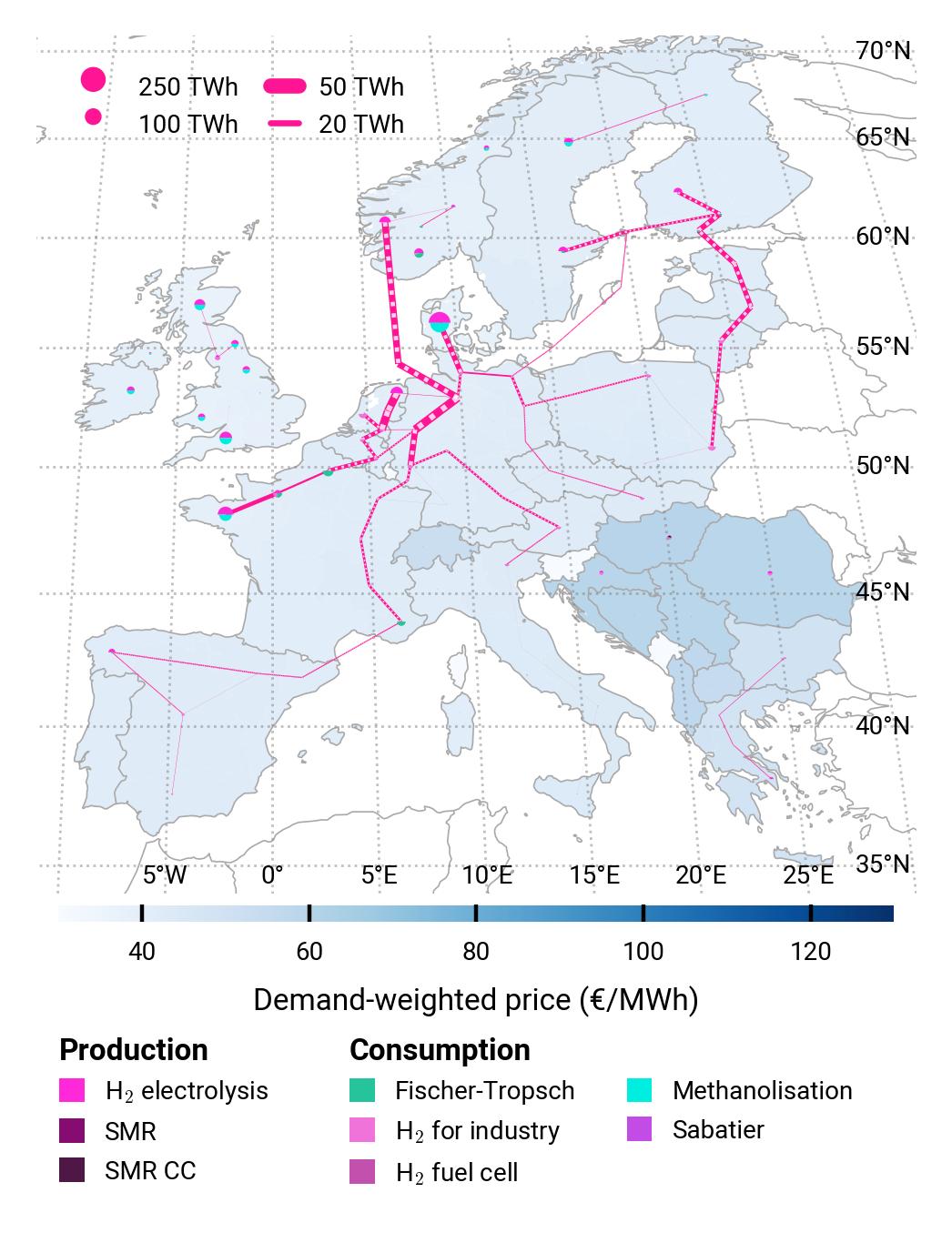}
      \caption{H$_2$ 2030.}
      \label{fig:PCI-n_lt_2030_h2}
  \end{subfigure}
  \begin{subfigure}[t]{0.32\textwidth}
      \vspace{0pt}
      \includegraphics[width=1\textwidth]{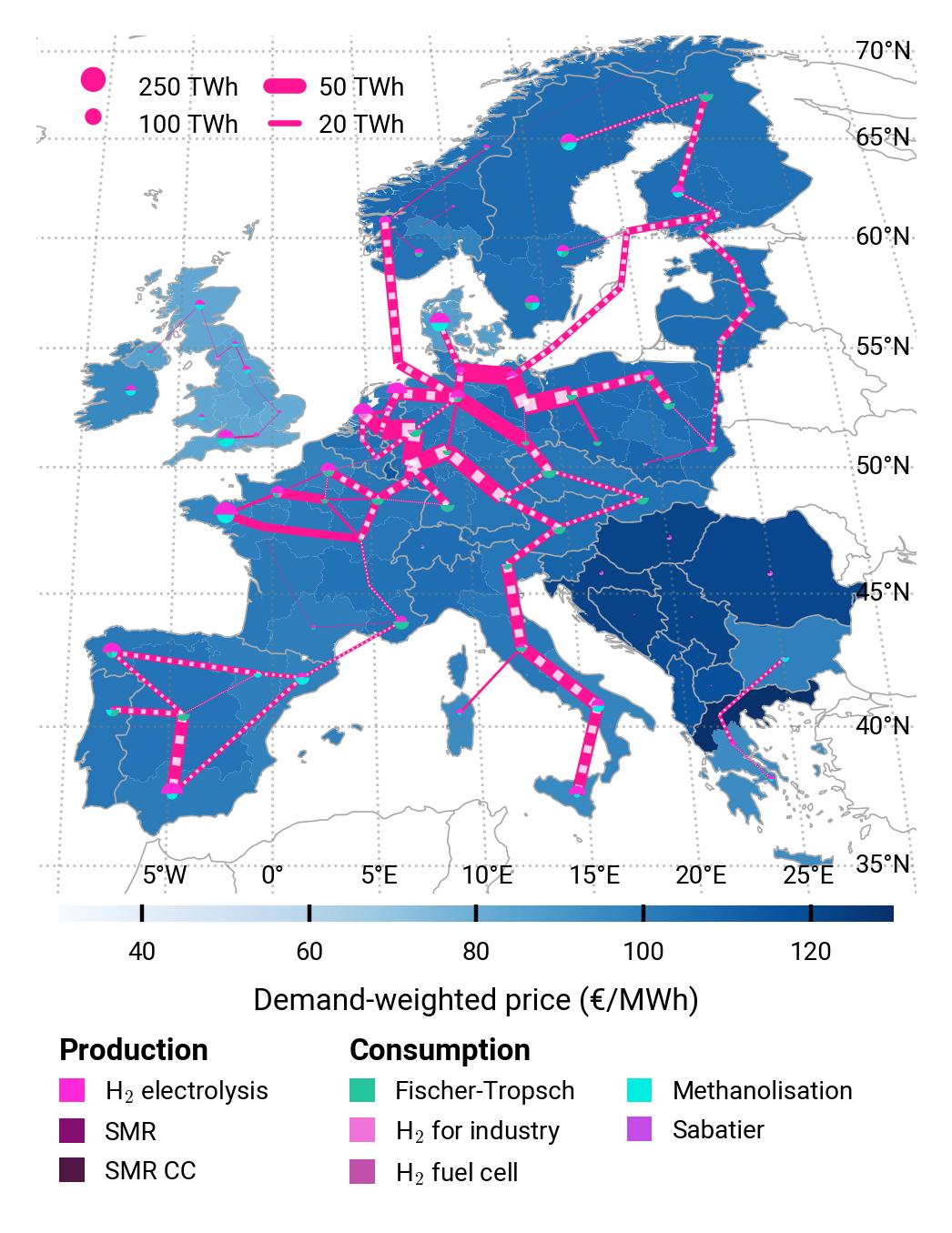}
      \caption{H$_2$ 2040.}
      \label{fig:PCI-n_lt_2040_h2}
  \end{subfigure}
  \begin{subfigure}[t]{0.32\textwidth}
    \vspace{0pt}
    \includegraphics[width=1\textwidth]{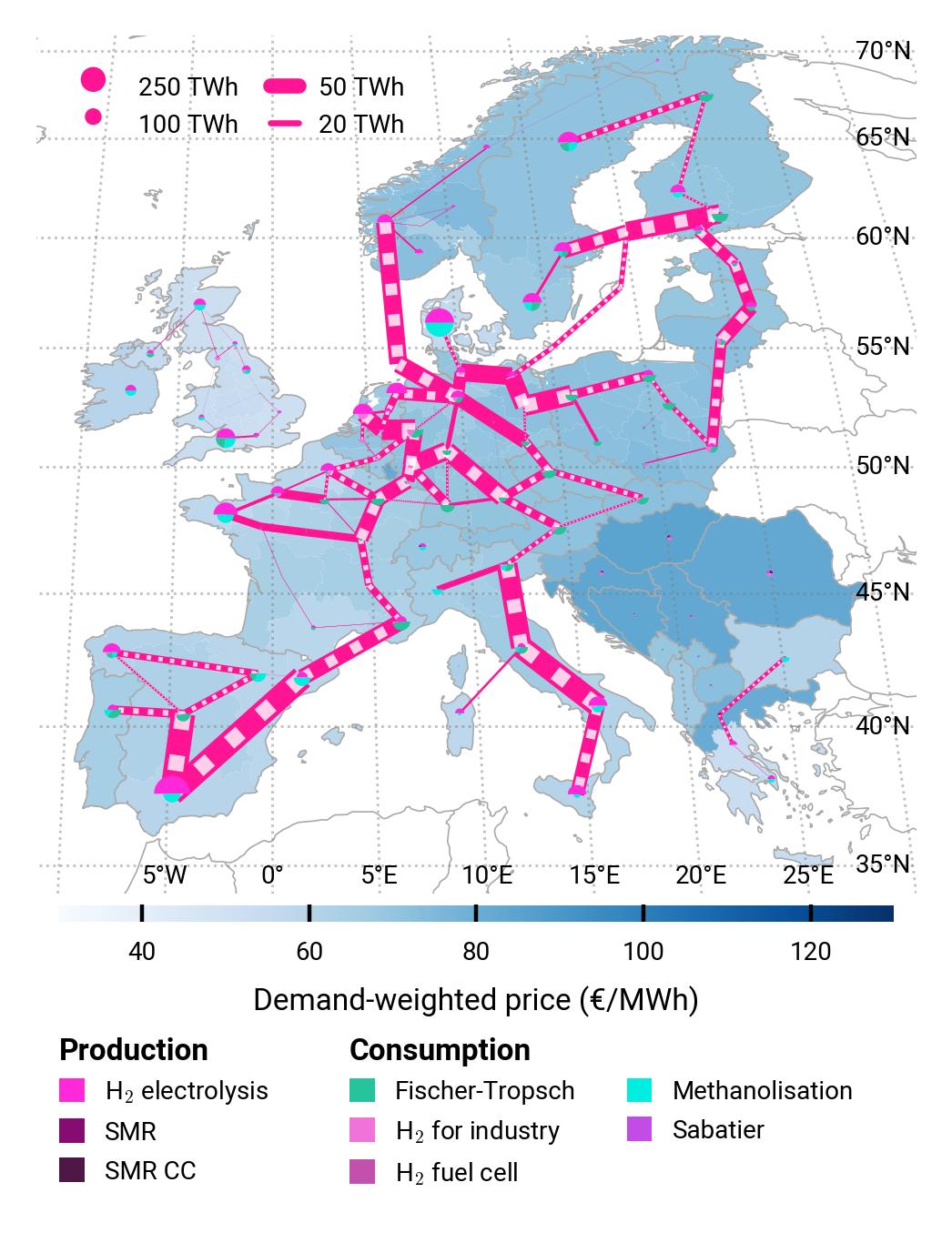}
    \caption{H$_2$ 2050.}
    \label{fig:PCI-n_lt_2050_h2}
  \end{subfigure}
  \begin{subfigure}[t]{0.32\textwidth}
      \vspace{0pt}
      \includegraphics[width=1\textwidth]{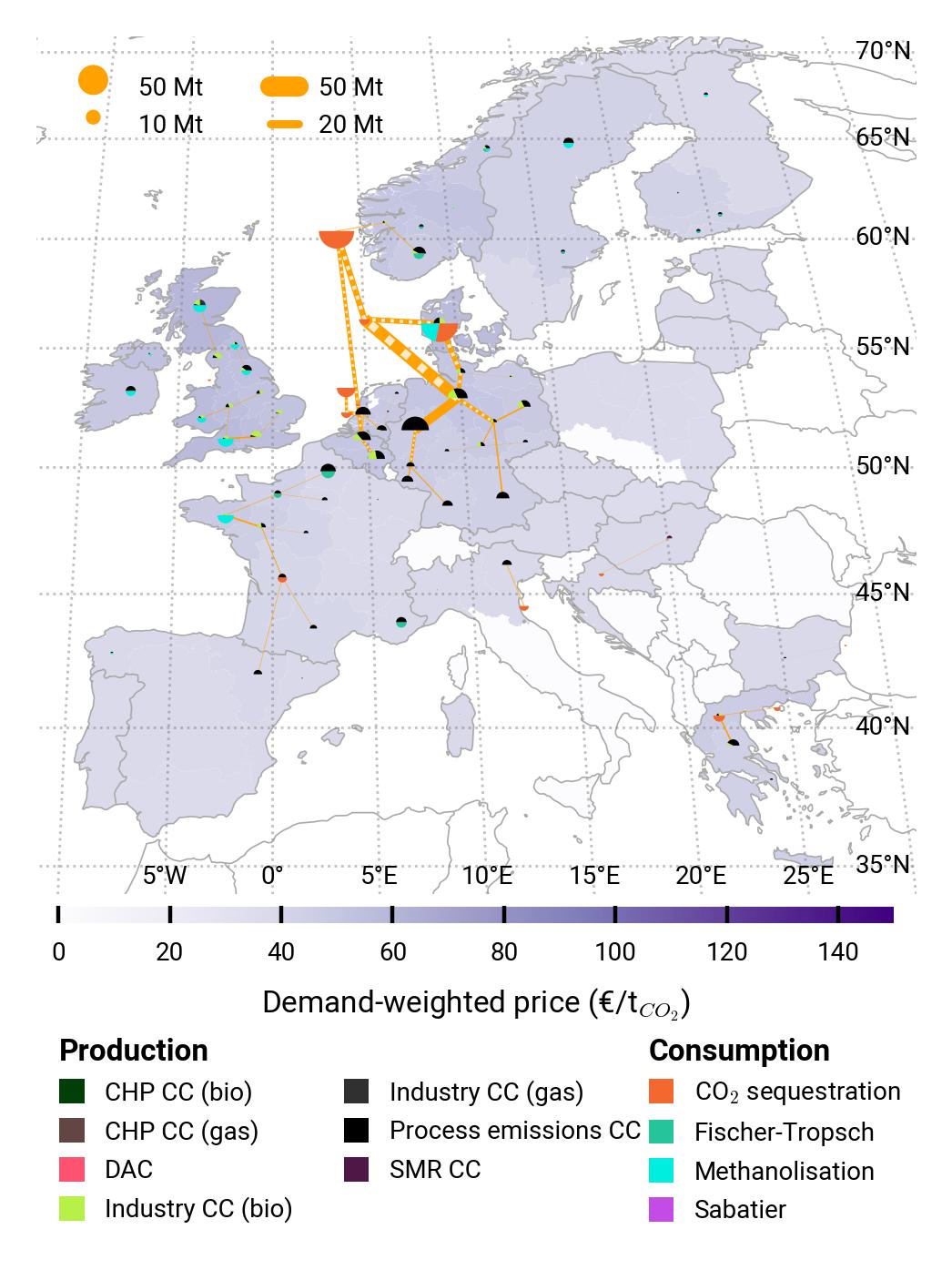} 
      \caption{CO$_2$ 2030.}
      \label{fig:PCI-n_lt_2030_co2}
  \end{subfigure}
  \begin{subfigure}[t]{0.32\textwidth}
      \vspace{0pt}
      \includegraphics[width=1\textwidth]{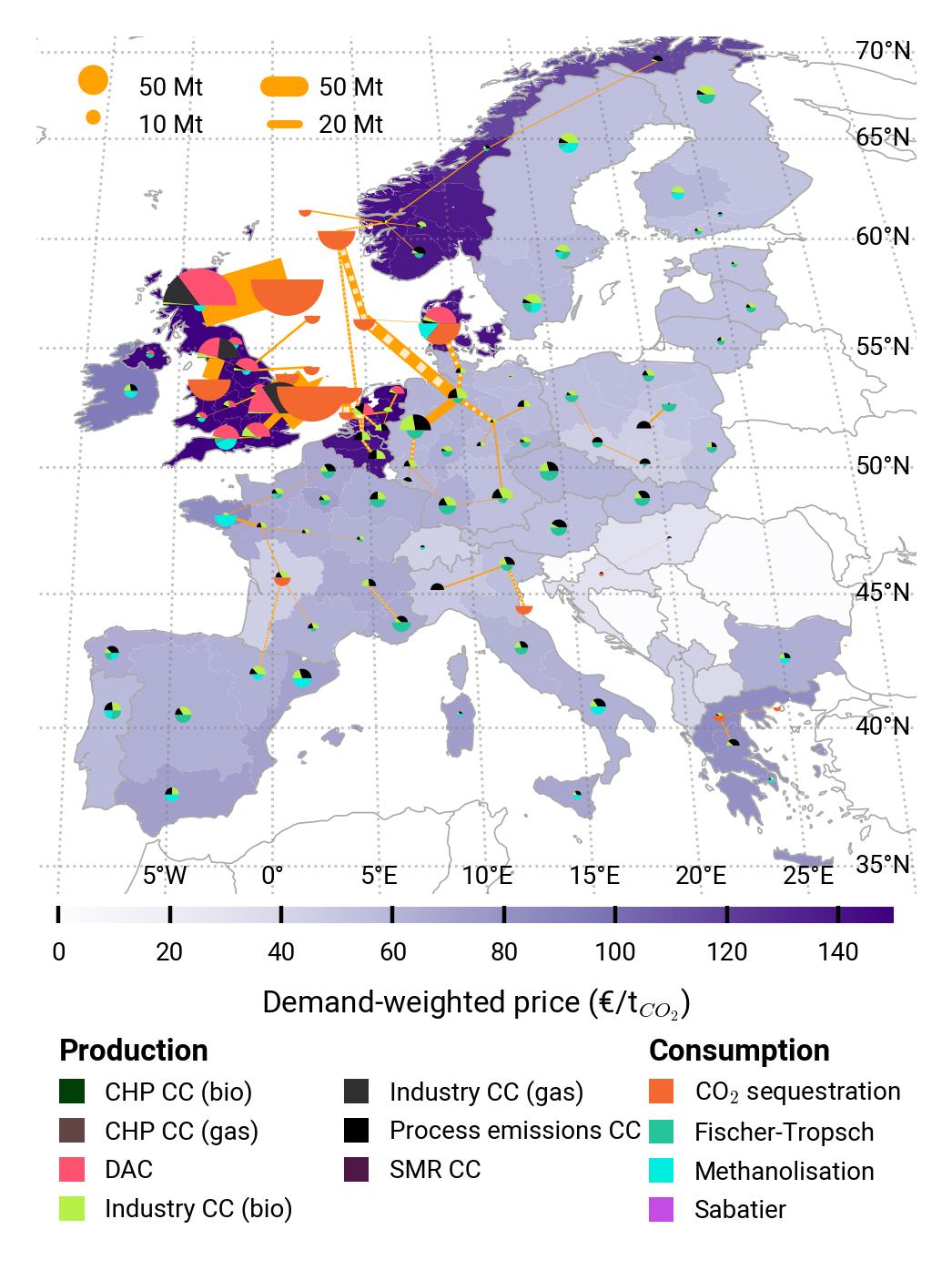} 
      \caption{CO$_2$ 2040.}
      \label{fig:PCI-n_lt_2040_co2}
  \end{subfigure}
  \begin{subfigure}[t]{0.32\textwidth}
      \vspace{0pt}
      \includegraphics[width=1\textwidth]{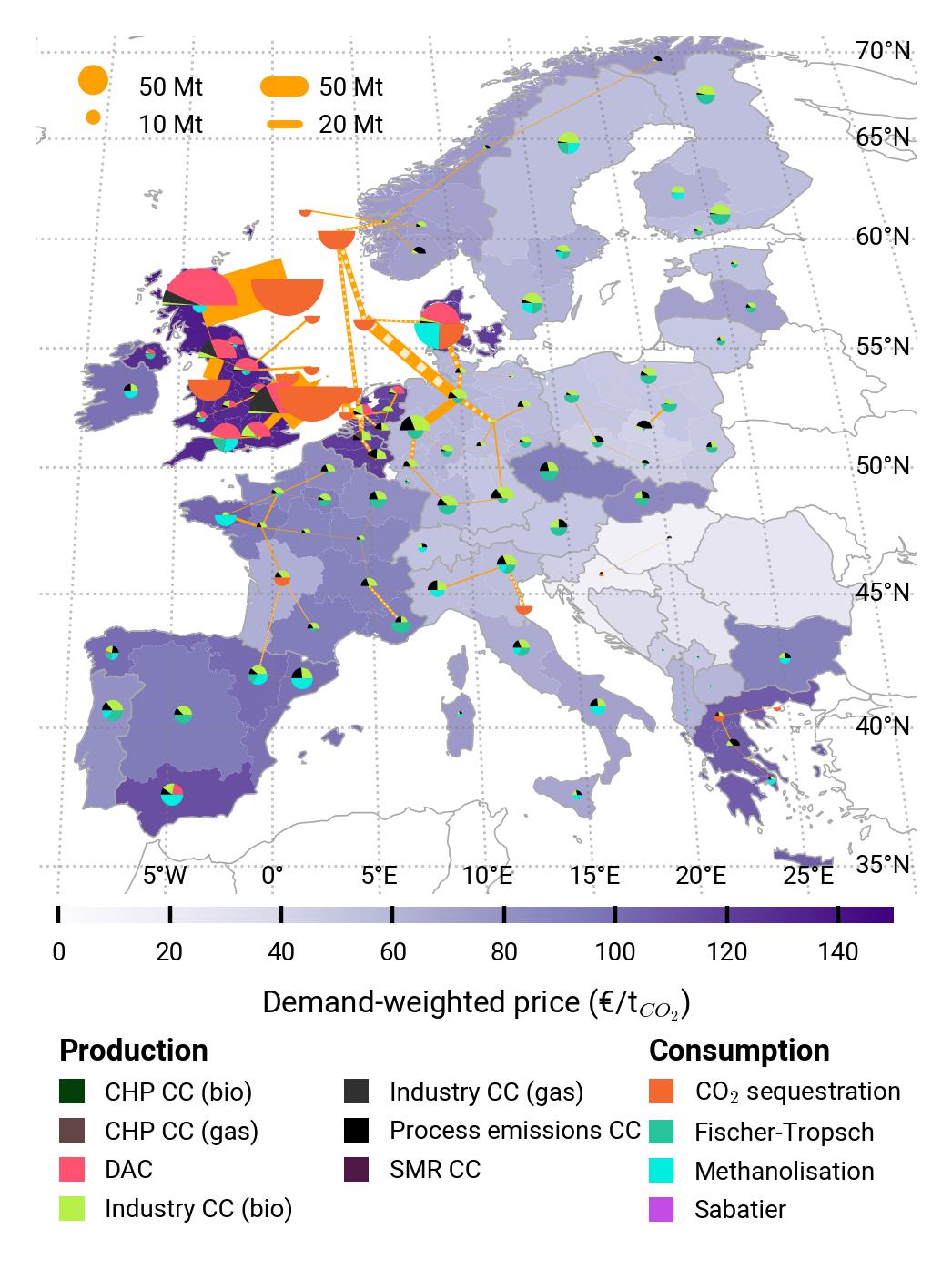} 
      \caption{CO$_2$ 2050.}
      \label{fig:PCI-n_lt_2050_co2}
  \end{subfigure}
  \vspace{0.3cm}
  \caption{\textit{PCI-PMI nat.} long-term scenario --- Regional distribution of H$_2$ and CO$_2$ production, utilisation, storage, transport and price. Note that both the H$_2$ and CO$_2$ price refer to their value as a commodity, i.e., price is higher where there is a demand for it.}
  \label{fig:PCI-n_lt}
\end{figure*}

\begin{figure*}[htbp]
  \centering
  \begin{subfigure}[t]{0.32\textwidth}
      \vspace{0pt}
      \includegraphics[width=1\textwidth]{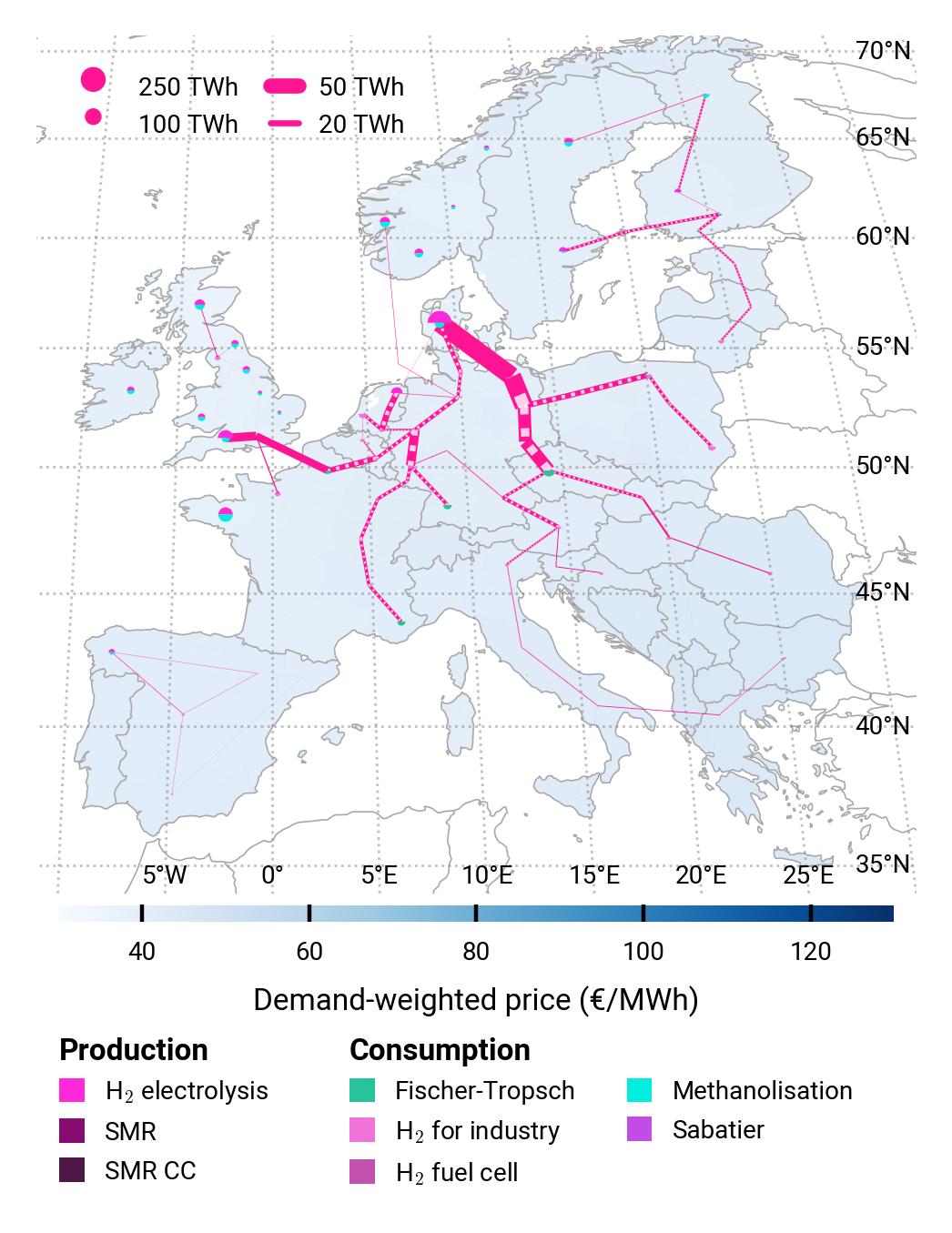}
      \caption{H$_2$ 2030.}
      \label{fig:PCI-in_lt_2030_h2}
  \end{subfigure}
  \begin{subfigure}[t]{0.32\textwidth}
      \vspace{0pt}
      \includegraphics[width=1\textwidth]{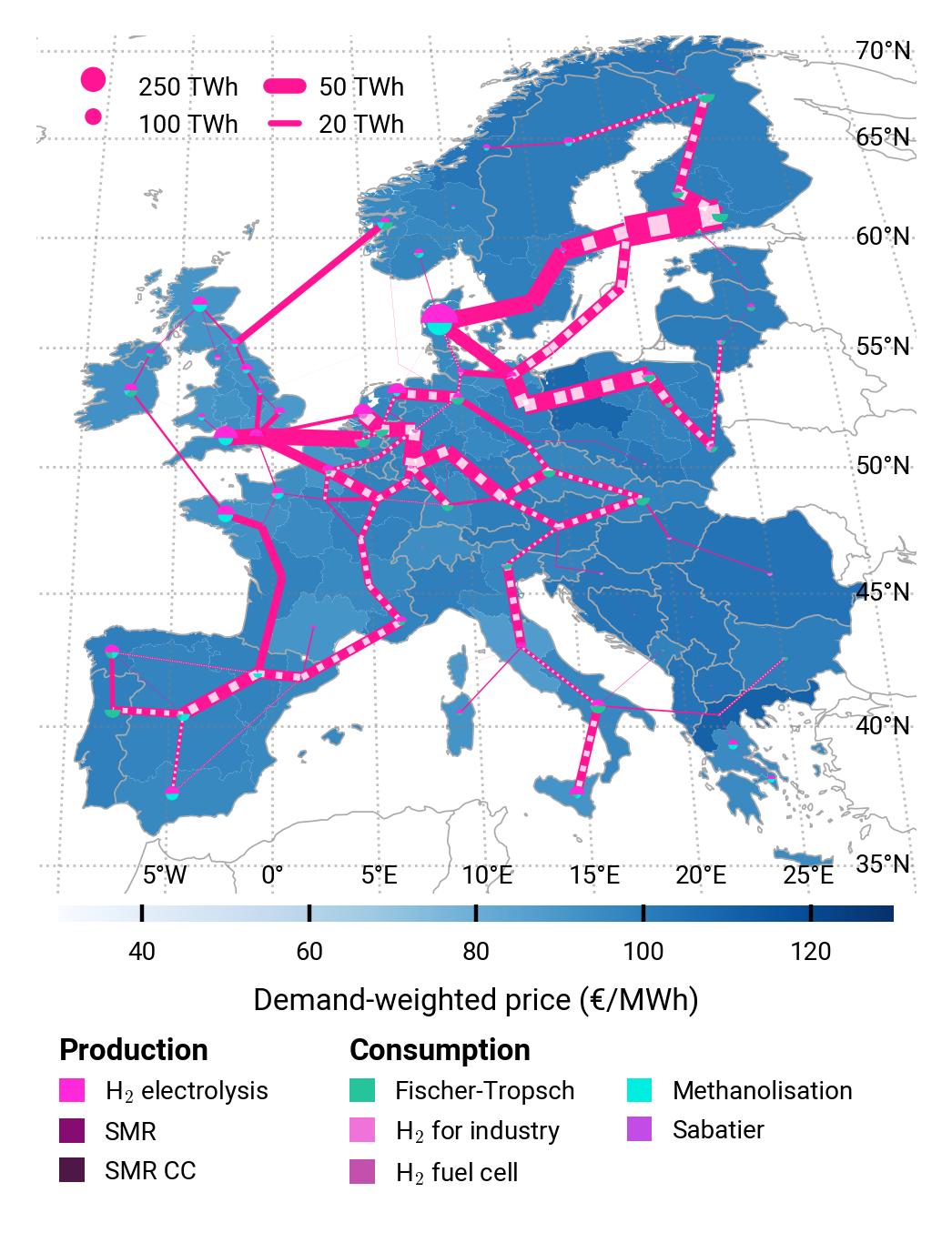}
      \caption{H$_2$ 2040.}
      \label{fig:PCI-in_lt_2040_h2}
  \end{subfigure}
  \begin{subfigure}[t]{0.32\textwidth}
    \vspace{0pt}
    \includegraphics[width=1\textwidth]{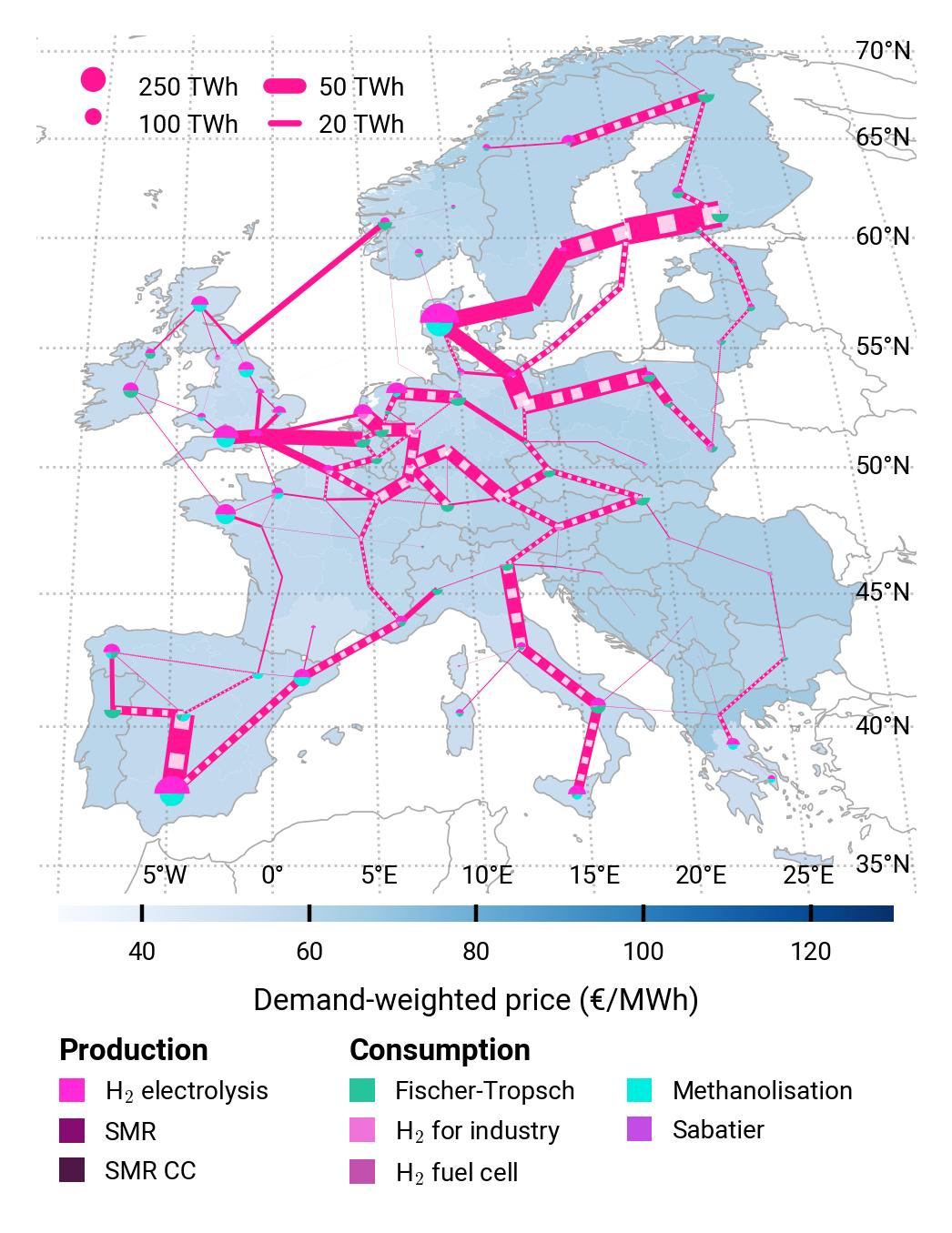}
    \caption{H$_2$ 2050.}
    \label{fig:PCI-in_lt_2050_h2}
  \end{subfigure}
  \begin{subfigure}[t]{0.32\textwidth}
      \vspace{0pt}
      \includegraphics[width=1\textwidth]{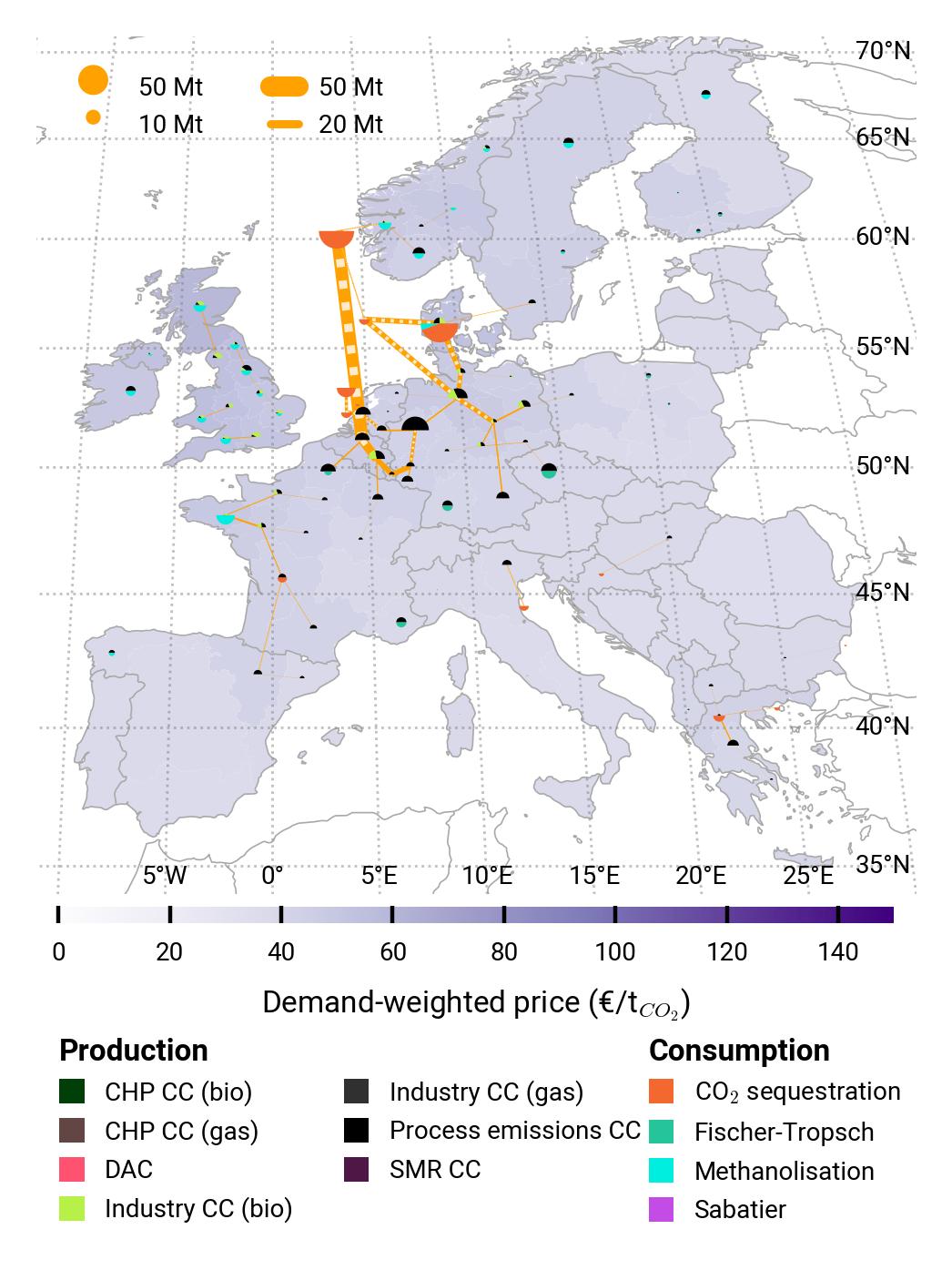} 
      \caption{CO$_2$ 2030.}
      \label{fig:PCI-in_lt_2030_co2}
  \end{subfigure}
  \begin{subfigure}[t]{0.32\textwidth}
      \vspace{0pt}
      \includegraphics[width=1\textwidth]{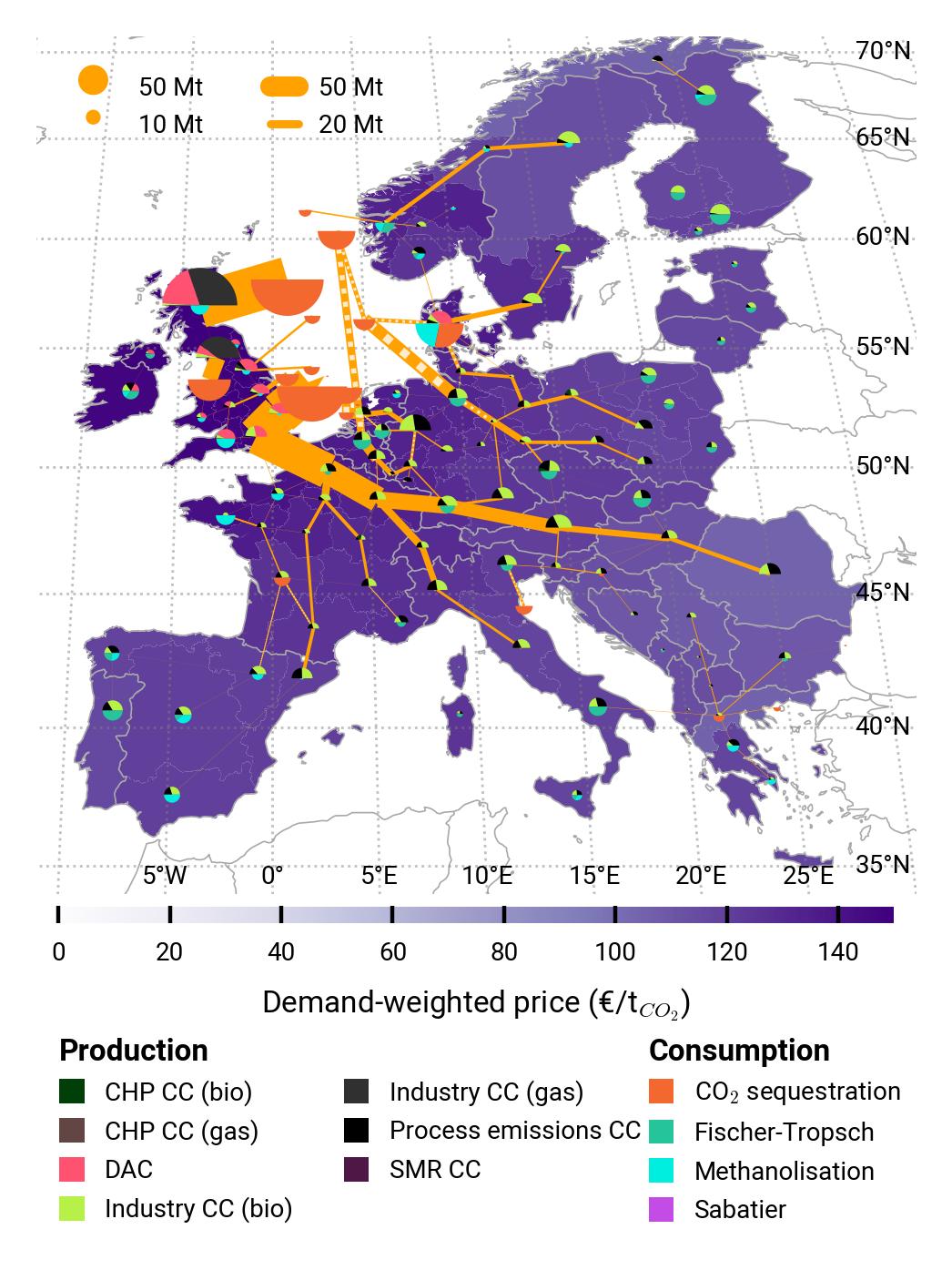} 
      \caption{CO$_2$ 2040.}
      \label{fig:PCI-in_lt_2040_co2}
  \end{subfigure}
  \begin{subfigure}[t]{0.32\textwidth}
      \vspace{0pt}
      \includegraphics[width=1\textwidth]{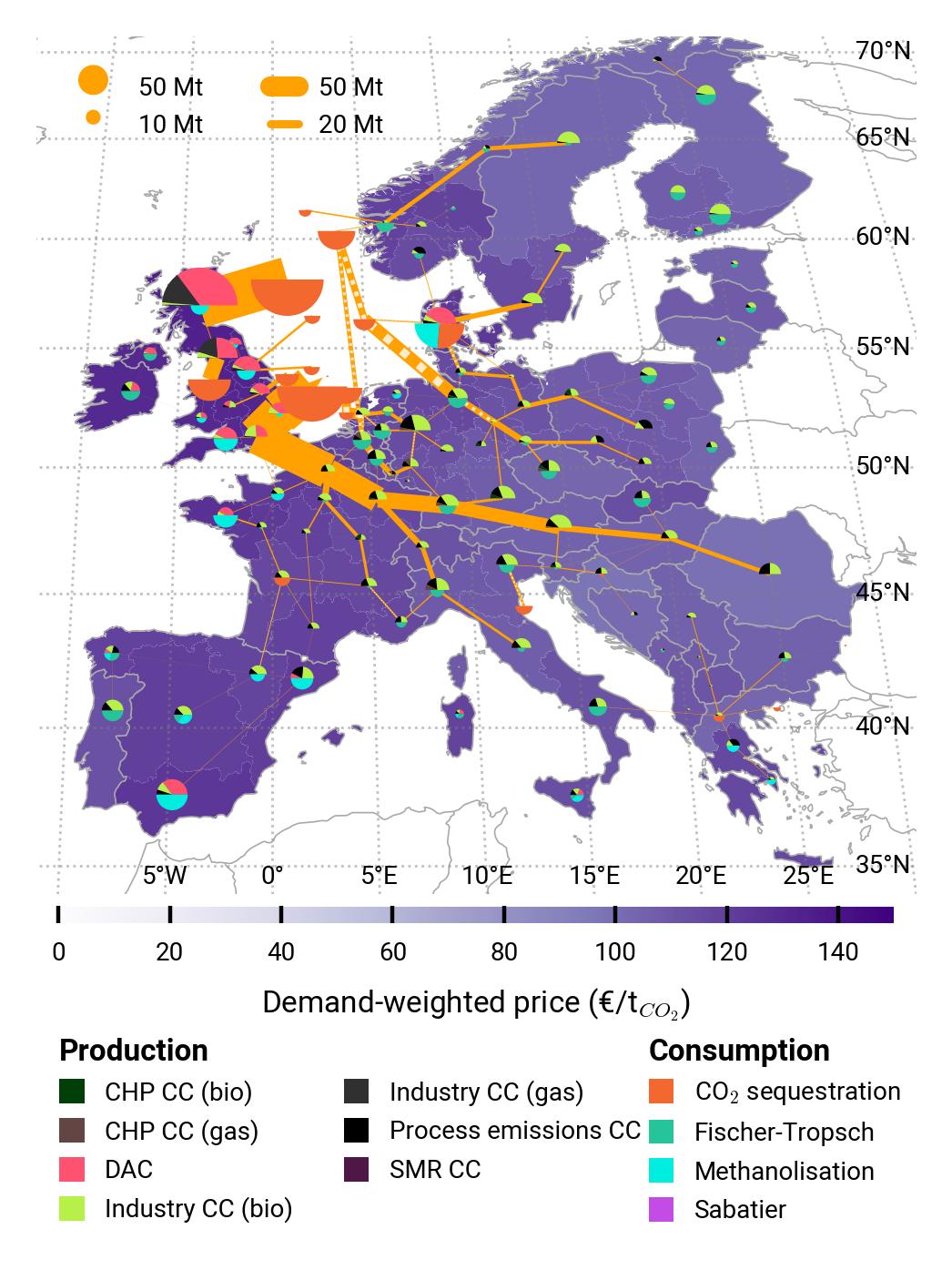} 
      \caption{CO$_2$ 2050.}
      \label{fig:PCI-in_lt_2050_co2}
  \end{subfigure}
  \vspace{0.3cm}
  \caption{\textit{PCI-PMI internat.} long-term scenario --- Regional distribution of H$_2$ and CO$_2$ production, utilisation, storage, transport and price. Note that both the H$_2$ and CO$_2$ price refer to their value as a commodity, i.e., price is higher where there is a demand for it.}
  \label{fig:PCI-in_lt}
\end{figure*}

\begin{figure*}[htbp]
  \centering
  \begin{subfigure}[t]{0.32\textwidth}
      \vspace{0pt}
      \includegraphics[width=1\textwidth]{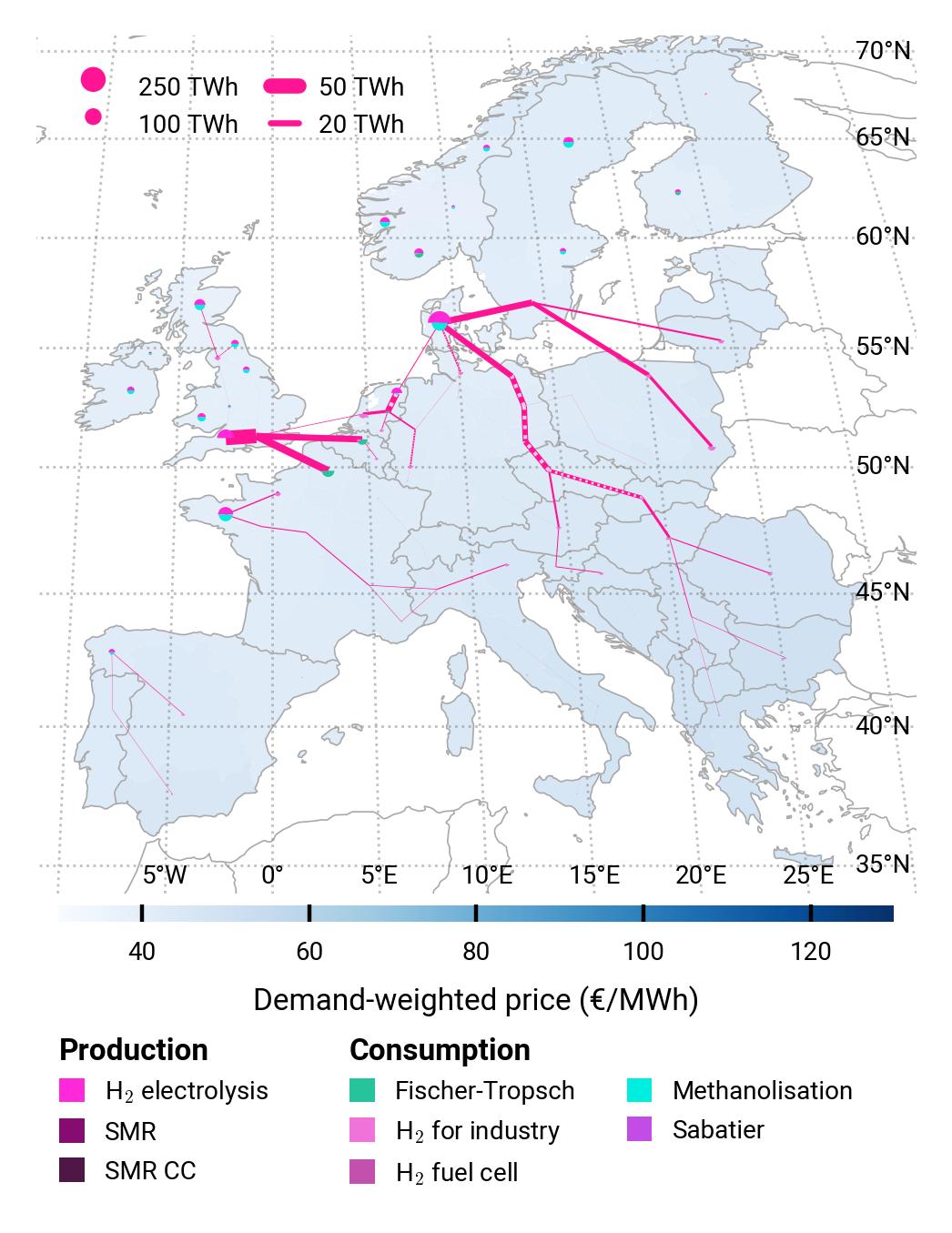}
      \caption{H$_2$ 2030.}
      \label{fig:CP_lt_2030_h2}
  \end{subfigure}
  \begin{subfigure}[t]{0.32\textwidth}
      \vspace{0pt}
      \includegraphics[width=1\textwidth]{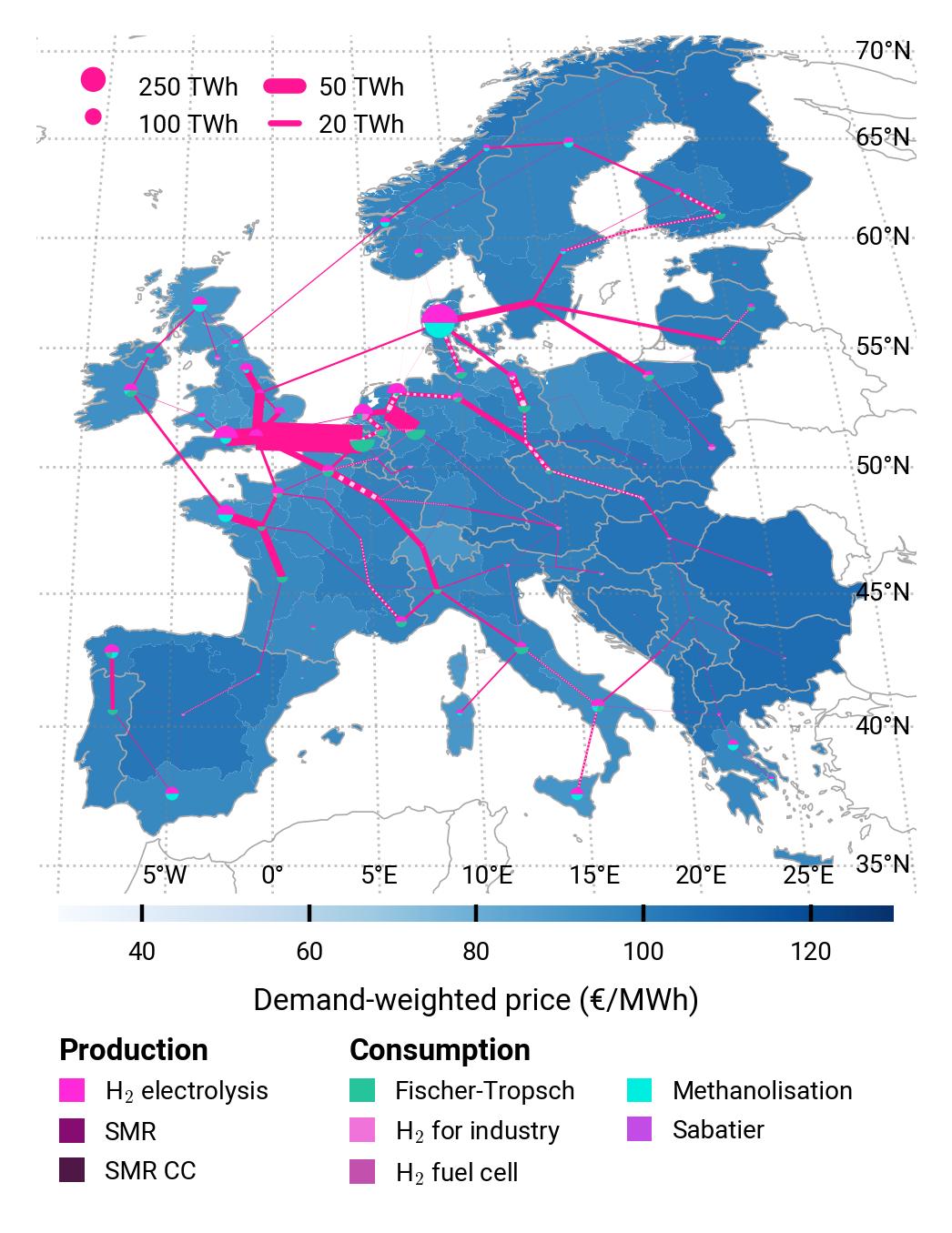}
      \caption{H$_2$ 2040.}
      \label{fig:CP_lt_2040_h2}
  \end{subfigure}
  \begin{subfigure}[t]{0.32\textwidth}
    \vspace{0pt}
    \includegraphics[width=1\textwidth]{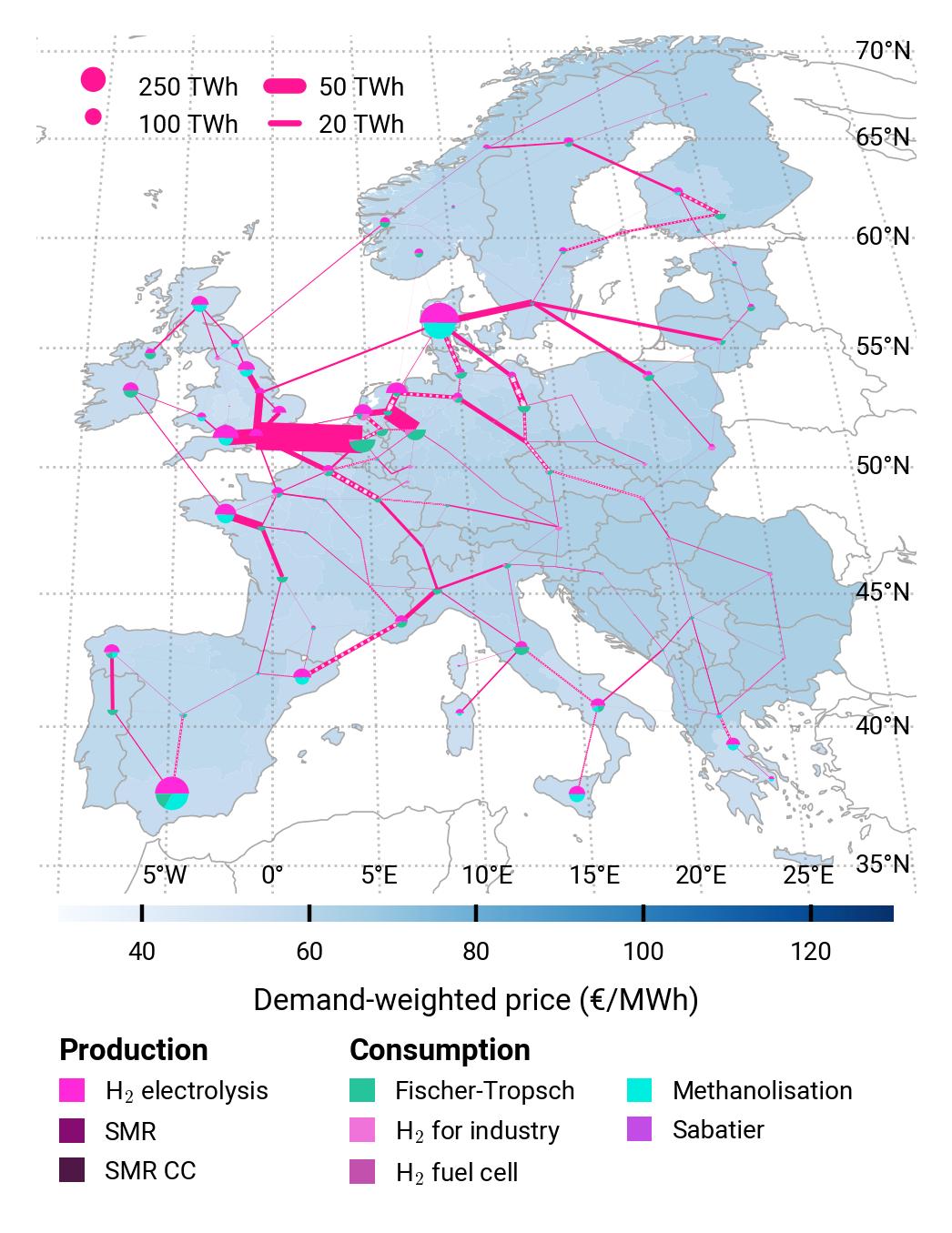}
    \caption{H$_2$ 2050.}
    \label{fig:CP_lt_2050_h2}
  \end{subfigure}
  \begin{subfigure}[t]{0.32\textwidth}
      \vspace{0pt}
      \includegraphics[width=1\textwidth]{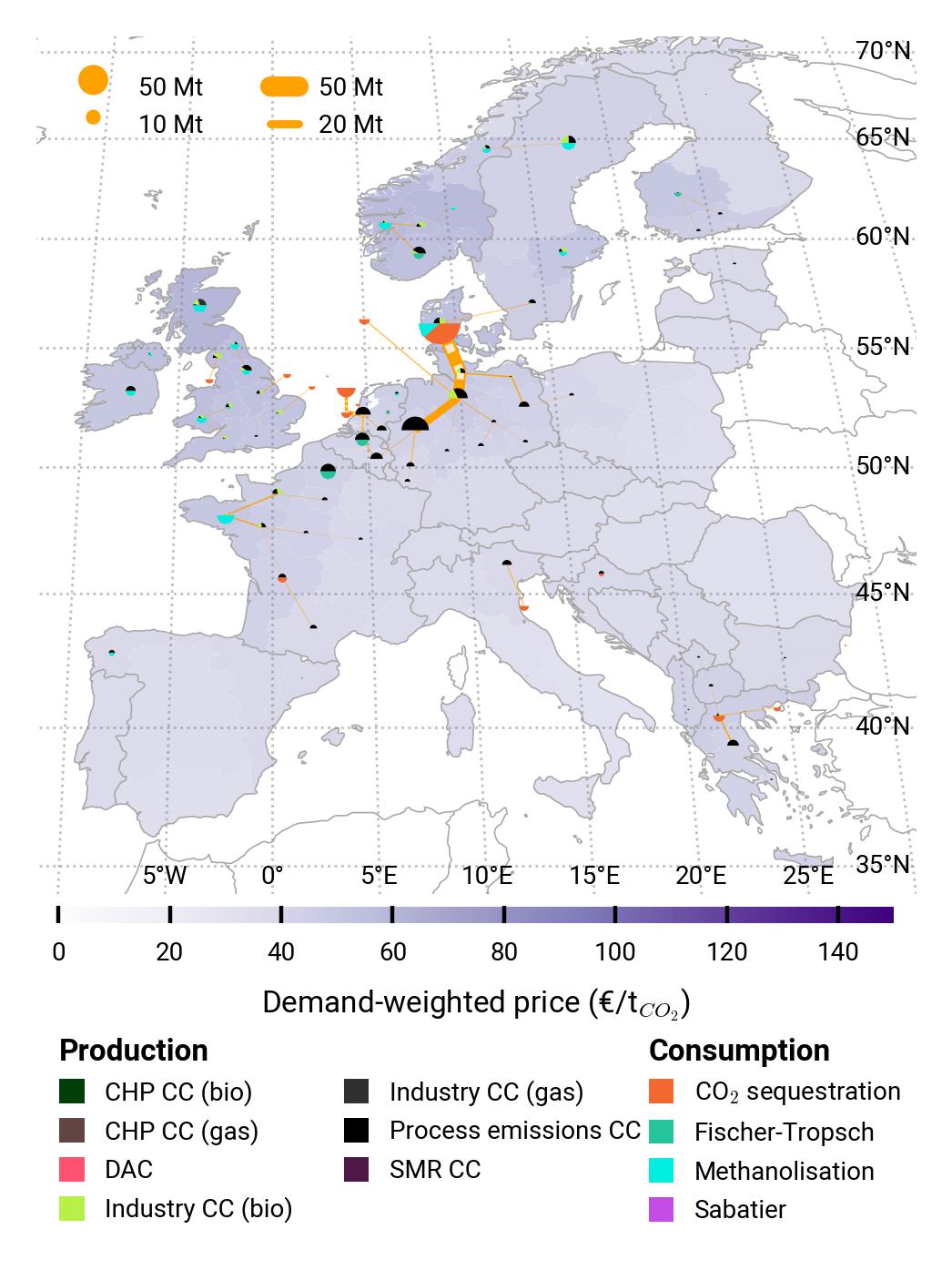} 
      \caption{CO$_2$ 2030.}
      \label{fig:CP_lt_2030_co2}
  \end{subfigure}
  \begin{subfigure}[t]{0.32\textwidth}
      \vspace{0pt}
      \includegraphics[width=1\textwidth]{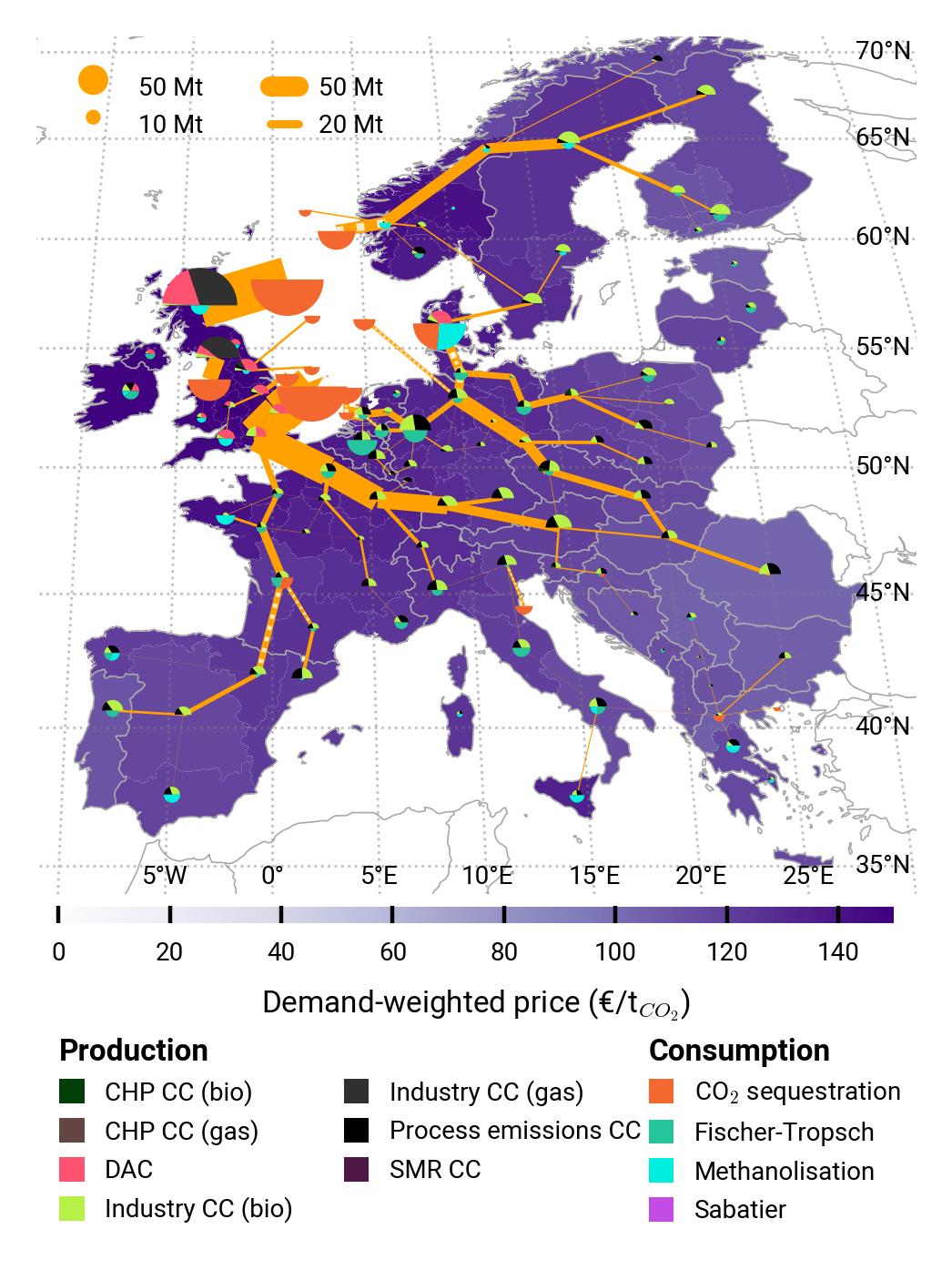} 
      \caption{CO$_2$ 2040.}
      \label{fig:CP_lt_2040_co2}
  \end{subfigure}
  \begin{subfigure}[t]{0.32\textwidth}
      \vspace{0pt}
      \includegraphics[width=1\textwidth]{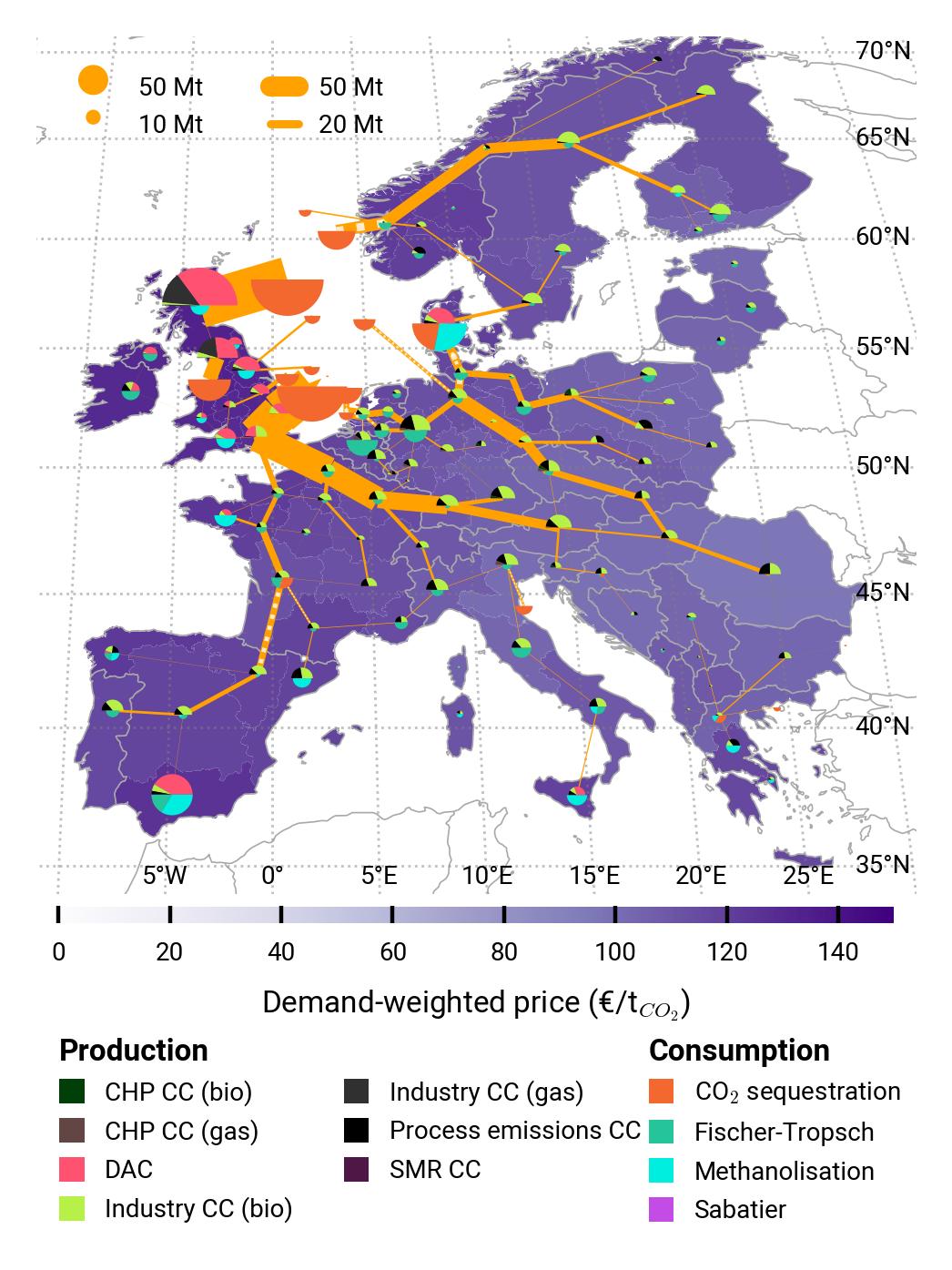} 
      \caption{CO$_2$ 2050.}
      \label{fig:CP_lt_2050_co2}
  \end{subfigure}
  \vspace{0.3cm}
  \caption{\textit{Central Planning} long-term scenario --- Regional distribution of H$_2$ and CO$_2$ production, utilisation, storage, transport and price. Note that both the H$_2$ and CO$_2$ price refer to their value as a commodity, i.e., price is higher where there is a demand for it.}
  \label{fig:CP_lt}
\end{figure*}

\end{appendices}


\clearpage
\bibliography{references}

\end{document}